\documentclass[10pt,journal,compsoc]{IEEEtran}

\ifCLASSOPTIONcompsoc%
  \usepackage[nocompress]{cite}
\else
  \usepackage{cite}
\fi

\ifCLASSINFOpdf%
  \usepackage[pdftex]{graphicx}
  \graphicspath{{../pdf/}{../jpeg/}}
  \DeclareGraphicsExtensions{.pdf,.jpeg,.png}
\else
  \usepackage[dvips]{graphicx}
  \graphicspath{{../eps/}}
  \DeclareGraphicsExtensions{.eps}
\fi
\usepackage{mwe}
\usepackage[dvipsnames]{xcolor}
\usepackage{xprintlen}
\usepackage{pgfplots}
\pgfplotsset{compat=1.16}
\usepackage{booktabs}
\usepackage{multirow}
\usepackage{tabularx}
\usepackage{adjustbox}
\usepackage{makecell}
\usepackage{rotating}
\usepackage{calc}

\usepackage{amsmath,amssymb,amsfonts}
\usepackage{algorithmic}
\usepackage[free-standing-units,per-mode=repeated-symbol,mode=text,
binary-units=true,detect-weight=true,detect-family=true]{siunitx}
\usepackage{textcomp}

\usepackage{array}
\usepackage{parcolumns}
\usepackage{xspace}
\usepackage{listings}
\usepackage{enumitem}

\usepackage{url}
\usepackage{glossaries}
\usepackage[hidelinks]{hyperref}
\hypersetup{breaklinks=true}
\usepackage[capitalise,nameinlink]{cleveref}

\usepackage{microtype}
\usepackage{soul}

\definecolor{color1}{HTML}{f94144}
\definecolor{color2}{HTML}{f9c74f}
\definecolor{color3}{HTML}{90be6d}
\definecolor{color4}{HTML}{43aa8b}
\definecolor{color5}{HTML}{577590}
\colorlet{colorAlert}{Red}

\newacronym{abi}{ABI}{application binary interface}
\newacronym{ace}{ACE}{AXI Coherent Extensions}
\newacronym{alu}{ALU}{arithmetic logic unit}
\newacronym{amba}{AMBA}{Advanced Microcontroller Bus Architecture}
\newacronym{amo}{AMO}{atomic memory operation}
\newacronym{aot}{AOT}{ahead-of-time}
\newacronym{apb}{APB}{Advanced Peripheral Bus}
\newacronym{api}{API}{application programming interface}
\newacronym{asic}{ASIC}{application-specific integrated circuit}
\newacronym{axi}{AXI}{Advanced eXtensible Interface}
\newacronym{bfs}{BFS}{breadth-first search}
\newacronym{blas}{BLAS}{Basic Linear Algebra Subprograms}
\newacronym{cas}{CAS}{compare-and-swap}
\newacronym{cmos}{CMOS}{complementary metal-oxide-semiconductor}
\newacronym{cnn}{CNN}{convolutional neural network}
\newacronym{cpu}{CPU}{central processing unit}
\newacronym{csr}{CSR}{control and status register}
\newacronym{dbt}{DBT}{dynamic binary translation}
\newacronym{dct}{DCT}{discrete cosine transform}
\newacronym{dlp}{DLP}{data level parallelism}
\newacronym{dma}{DMA}{direct memory access}
\newacronym{dram}{DRAM}{dynamic random-access memory}
\newacronym{dsl}{DSL}{domain-specific language}
\newacronym{dsp}{DSP}{digital signal processing}
\newacronym{elf}{ELF}{Executable and Linkable Format}
\newacronym{fdsoi}{FD-SOI}{fully depleted silicon-on-insulator}
\newacronym{fpga}{FPGA}{field-programmable gate array}
\newacronym{fpu}{FPU}{floating-point unit}
\newacronym[longplural={general-purpose \acrlongpl{gpu}}]{gpgpu}{GPGPU}{general-purpose \gls{gpu}}
\newacronym{gpu}{GPU}{graphics processing unit}
\newacronym{hart}{hart}{hardware thread}
\newacronym{hbm}{HBM}{High Bandwidth Memory}
\newacronym{hdl}{HDL}{hardware description language}
\newacronym{hero}{HERO}{Heterogeneous Embedded Research Platform}
\newacronym{hpc}{HPC}{high-performance computing}
\newacronym{ilp}{ILP}{instruction level parallelism}
\newacronym{ipc}{IPC}{instructions per cycle}
\newacronym{iot}{IoT}{Internet of Things}
\newacronym{ipu}{IPU}{integer processing unit}
\newacronym{ir}{IR}{intermediate representation}
\newacronym{isa}{ISA}{instruction set architecture}
\newacronym{issr}{ISSR}{indirection stream semantic register}
\newacronym{jit}{JIT}{just-in-time}
\newacronym{llc}{LLC}{last-level cache}
\newacronym{lrsc}{LRSC}{load-reserved/store-conditional}
\newacronym{lr}{LR}{load-reserved}
\newacronym{lsu}{LSU}{load-store unit}
\newacronym{mac}{MAC}{multiply–accumulate}
\newacronym{mimd}{MIMD}{multiple instruction, multiple data}
\newacronym{mmu}{MMU}{memory management unit}
\newacronym[longplural={networks-on-chip}]{noc}{NoC}{network-on-chip}
\newacronym{nuca}{NUCA}{non-uniform cache architecture}
\newacronym{numa}{NUMA}{non-uniform memory access}
\newacronym{pc}{PC}{program counter}
\newacronym{pe}{PE}{processing element}
\newacronym{pl}{PL}{programmable logic}
\newacronym{pmca}{PMCA}{programmable manycore accelerator}
\newacronym{pulp}{PULP}{Parallel Ultra Low Power}
\newacronym{raw}{RAW}{read-after-write}
\newacronym{rom}{ROM}{read-only memory}
\newacronym{rmw}{RMW}{read–modify–write}
\newacronym{rob}{ROB}{reorder buffer}
\newacronym{rvwmo}{RVWMO}{RISC-V Weak Memory Ordering}
\newacronym{ro}{RO}{read-only}
\newacronym{rtl}{RTL}{register-transfer level}
\newacronym{sbt}{SBT}{static binary translation}
\newacronym{scm}{SCM}{standard cell memory}
\newacronym{sc}{SC}{store-conditional}
\newacronym{sdf}{SDF}{Standard Delay Format}
\newacronym{simd}{SIMD}{single instruction, multiple data}
\newacronym{simt}{SIMT}{single instruction, multiple thread}
\newacronym{sm}{SM}{streaming multiprocessor}
\newacronym{soc}{SoC}{system-on-chip}
\newacronym[longplural={scratchpad memories}]{spm}{SPM}{scratchpad memory}
\newacronym{sram}{SRAM}{static random-access memory}
\newacronym{ssa}{SSA}{static single assignment}
\newacronym{ssr}{SSR}{stream semantic register}
\newacronym{tcdm}{TCDM}{tightly-coupled data memory}
\newacronym{tlp}{TLP}{thread-level parallelism}
\newacronym{vpu}{VPU}{vector processing unit}
\newacronym{vliw}{VLIW}{very long instruction word}
\newacronym{vnb}{VNB}{von Neumann bottleneck}
\newacronym{war}{WAR}{write-after-read}
\newacronym{waw}{WAW}{write-after-write}

\newcounter{undefinedreferences}
\newcommand{\checkreferences}{
	\ifnum\value{undefinedreferences} > 0
	\begin{center}
		\immediate\write18{wget -O Figures/protester.png -nc http://imgs.xkcd.com/comics/wikipedian_protester.png}
		\includegraphics[width=\textwidth]{protester.png}
	\end{center}
	\else
	No undefined references. Good!
	\fi
}

\DeclareSIUnit\core{core}
\DeclareSIUnit\tile{tile}
\DeclareSIUnit\request{req}
\DeclareSIUnit\cycle{cycle}
\DeclareSIUnit\erlang{E}
\DeclareSIUnit\flop{FLOP}
\DeclareSIUnit\flops{FLOPS}
\DeclareSIUnit\gate{GE}
\DeclareSIUnit\ge{GE}
\DeclareSIUnit\op{OP}
\DeclareSIUnit\ops{OPS}
\DeclareSIUnit\bps{bps}
\DeclareSIUnit\Bps{Bps}
\DeclareSIUnit\ipc{IPC}
\DeclareSIUnit[number-unit-product = ]\percent{\%}

\newcommand{\SIadj}[2]{\SI[number-unit-product={\text{-}}]{#1}{#2}}

\PackageWarning{Citation missing:}{#1!}

\PackageWarning{TODO:}{#1!}

\newcommand\riscv{RISC-V}
\newcommand\mempool{Mem\-Pool}

\newcommand\topology[1]{Top\ensuremath{_{\text{#1}}}}
\newcommand\by[2]{#1$\times$#2}
\newcommand\changed[1]{{#1}}
\newcommand\smallkernel{\emph{small}}
\newcommand\bigkernel{\emph{big}}
\newcommand\matmul{\emph{matmul}}
\newcommand\conv{\emph{2dconv}}
\newcommand\dct{\emph{dct}}
\newcommand\dotp{\emph{dotp}}
\newcommand\axpy{\emph{axpy}}
\newcommand\histequal{\emph{histogram equalization}}
\newcommand\raytracing{\emph{ray tracing}}
\newcommand\bfs{\emph{\acrlong{bfs}}}

\begin{document}
\bstctlcite{IEEEexample:BSTcontrol}
\title{MemPool: A Scalable Manycore Architecture with a Low-Latency Shared L1 Memory}

\author{Samuel~Riedel, Matheus~Cavalcante, Renzo~Andri, and~Luca~Benini% stops a space
  \IEEEcompsocitemizethanks{%
    \IEEEcompsocthanksitem{} Samuel~Riedel, Matheus~Cavalcante, and Luca~Benini are with the
     Integrated Systems Laboratory (IIS), ETH Zurich, Switzerland. E-mail:
     \{sriedel, matheus, lbenini\}@iis.ee.ethz.ch% stops an unwanted space
    \IEEEcompsocthanksitem{} Renzo~Andri is an Independent Researcher, 8050 Zurich, Switzerland. E-mail: info@renzo.ch% stops an unwanted space
    \IEEEcompsocthanksitem{} Luca Benini is also with the Department of Electrical, Electronic and
     Information Engineering (DEI), University of Bologna,
     Italy.
    \IEEEcompsocthanksitem{IEEE Publication DOI: \href{https://doi.org/10.1109/TC.2023.3307796}{10.1109/TC.2023.3307796}}
  }% stops an unwanted space
}

% Copyright
\IEEEoverridecommandlockouts
\IEEEpubid{
  \begin{minipage}[c]{\textwidth}
    \centering
    \tiny
    \copyright~2023 IEEE. Personal use of this material is permitted. Permission from IEEE must be obtained for all other uses, in any current or future media, including reprinting/republishing this material \\ for advertising or promotional purposes, creating new collective works, for resale or redistribution to servers or lists, or reuse of any copyrighted component of this work in other works.\hfill
  \end{minipage}
}

\markboth{}%
{Riedel \MakeLowercase{\textit{et al.}}: MemPool}

\IEEEtitleabstractindextext{%

\begin{abstract}
\glsresetall
\glsunset{pe}
\glsunset{dma}
\glsunset{gpgpu}
Shared L1 memory clusters are a common architectural pattern (e.g., in \glspl{gpgpu}) for building efficient and flexible multi-processing-element (PE) engines. However, it is a common belief that these tightly-coupled clusters would not scale beyond a few tens of \glspl{pe}. In this work, we tackle scaling shared L1 clusters to hundreds of \glspl{pe} while supporting a flexible and productive programming model and maintaining high efficiency. We present \mempool{}, a manycore system with 256 RV32IMAXpulpimg ``Snitch'' cores featuring application-tunable functional units. We designed and implemented an efficient low-latency \gls{pe} to L1-memory interconnect, an optimized instruction path to ensure each \gls{pe}'s independent execution, and a powerful \gls{dma} engine and system interconnect to stream data in and out. \mempool{} is easy to program, with all the cores sharing a global view of a large, multi-banked, L1 scratchpad memory, accessible within at most five cycles in the absence of conflicts. We provide multiple runtimes to program \mempool{} at different abstraction levels and illustrate its versatility with a wide set of applications. \mempool{} runs at \SI{600}{\mega\hertz} (60 gate delays) in typical conditions (TT/\SI{0.80}{\volt}\kern-.1em/\SI{25}{\celsius}) in 22\,nm FDX technology and achieves a performance of up to \SI{229}{\giga\ops} or \SI{180}{\giga\ops\per\watt} with less than \SI{2}{\percent} of execution stalls.
\end{abstract}

% Note that keywords are not normally used for peer-reviewed papers.
\begin{IEEEkeywords}
%TODO: They have to be ordered alphabetically
Manycore, RISC-V, general-purpose
\end{IEEEkeywords}
}

% Make the title area
\maketitle

% Print the copyright
\IEEEpubidadjcol

\glsresetall

\section{Introduction}\label{sec:introduction}

\IEEEPARstart{M}{ulticore} systems are essential to efficiently run today's compute-intensive, extremely parallel workloads ranging from genomics over computational photography and machine learning to graph processing~\cite{Muralidhar2022,Li2010}. The pursuit of energy-efficient parallel processing has led to a continuous increase in core count in modern computer architectures~\cite{Karlrupp2022}. For example, NVIDIA's H100 Tensor Core \glspl{gpu} feature 144 \glspl{sm} with tens of \glspl{pe} each~\cite{NVIDIACorporation2022}, while machine learning accelerators like Cerebras' Wafer-Scale Engine boast up to \num{850000} cores~\cite{Rocki2020}.

Multi-core systems span a broad spectrum from general-purpose to highly domain-specific. In the domain-specific corner, we find specialized systolic-array-based architectures, such as Google's Pixel Visual Core~\cite{Redgrave2018} or Tensor Processing Unit~\cite{Hennessy2017}. They scale to hundreds of cores, thanks to their neighbor-to-neighbor communication. While those systems are well suited for their target kernels, their rigid, specialized interconnect and execution scheme restrict their application domain. Moreover, their programming model is notoriously difficult to manage. In contrast, general-purpose processors such as Apple's M1 Ultra~\cite{Apple2022}, Intel's Core-i9~\cite{IntelCorporation2022}, and Ampere's Altra~\cite{Ampere2022} combine a few high-performance cores sharing large caches. Their complex cores and caches are versatile and facilitate programming but impose a power cost and limit scalability beyond a few tens of cores.

A typical architectural pattern that aims for an optimal compromise between efficiency and flexibility is a tightly coupled cluster of simple cores sharing an L1 data memory through a low-latency interconnect. We can find instances of this architectural pattern across numerous domains, from \glspl{sm} of \glspl{gpu}~\cite{NVIDIACorporation2022}, to the ultra-low-power GreenWaves' GAP9 processor~\cite{GreenWavesTechnologies2021}, to the high-performance embedded signal processing with the Kalray processor clusters~\cite{DupontdeDinechin2021}, to aerospace applications with the Ramon Chips' RC64 system~\cite{Ginosar2016}. The compute cluster approach can efficiently solve a wide range of today's computing problems. However, it is commonly understood that scaling such tightly coupled clusters to more than a few tens of cores is unfeasible due to the low-latency access requirements to the L1 memory, the aggregate instruction stream's bandwidth of all \glspl{pe}, and the increasing complexity of interconnects.

Computer architects have devised ways to push performance for tightly coupled clusters. For example, \glspl{sm} in \glspl{gpgpu} implement a \gls{simt} scheme, replicating functional units that work in lock-step and sacrifice instruction stream diversity across individual cores. A different approach to scaling beyond tens of cores is replicating the simple compute cluster. This approach is found in \glspl{gpu} as well as high-performance accelerators such as Manticore~\cite{Zaruba2021} or Esperanto's ET-SoC-1~\cite{Ditzel2021}. However, multiple clusters with their L1 memories expand the memory hierarchy and introduce communication overhead between them. Furthermore, the additional hierarchy complicates parallelization and the programming model.

This paper tackles the challenge of building an energy-efficient, general-purpose manycore system by scaling the versatile single cluster architecture to hundreds of cores. We set out to design a manycore system that avoids the problems related to \gls{simt} or multi-cluster designs. Our system has independent and individually programmable, efficient, small cores with low-latency access to a globally shared L1 memory. With efficient synchronization techniques and fast \gls{dma} engines to move data in and out of the cluster's L1 memory, we aim to design a general-purpose manycore system with a streamlined programming model.

We present \mempool{}, an open-source, parametric, and flexible manycore architecture scalable to hundreds of independently programmable 32-bit \riscv{} cores sharing a large L1 \gls{spm}. Specifically, we focus on a maximal \mempool{} configuration with \num{256} cores and \SI{1}{\mebi\byte} of L1 \gls{spm} accessible in at most five cycles in the absence of contention. By leveraging small cores designed to tolerate outstanding memory instructions, we can hide L1 interconnect latency and close the gap with respect to the performance of an ideally scaled, physically not implementable, single-cycle access L1 cluster. A highly optimized, scalable, cache-based instruction path supplies cores with instructions, while a \gls{dma} engine custom-designed for \mempool{} efficiently moves data in and out of the L1 \gls{spm}.

With compiler and runtime support to take full advantage of \mempool{}'s architecture, we present a single-cluster manycore architecture that can easily be programmed and efficiently runs not only data-intensive image processing algorithms with regular and predictable memory access patterns but also workloads like graph processing which are characterized by irregular and unpredictable L1 accesses.

This paper extends our earlier work~\cite{Cavalcante2021} and presents the following contributions:
\begin{itemize}
  \item The \mempool{} architecture; a flexible manycore architecture with shared L1 \gls{spm} (\cref{sec:architecture_overview}). Specifically:
  \begin{itemize}
    \item A physical-aware design of the low-latency scalable L1 data memory interconnect combined with a lightweight and transparent memory addressing scheme that keeps the memory region that is most often accessed by a core in the same memory bank or close by with minimal access latency and energy consumption (\cref{sec:l1_memory_interconnect});
    \item An instruction cache architecture highly optimized for energy-efficiency (\cref{sec:instruction_cache});
    \item A scalable hierarchical system interconnect, containing a cache hierarchy and a specialized distributed \gls{dma} engine to feed the full \mempool{} cluster with instructions and data (\cref{sec:system_interconnect});
  \end{itemize}
  \item The full physical implementation and performance, power, and area analysis of \mempool{} in GlobalFoundries 22FDX \gls{fdsoi} technology node (\cref{sec:implementation});
  \item Compiler support and software runtimes to efficiently program \mempool{}, including an implementation of the OpenMP standard and support for a \gls{dsl}, namely Halide~\cite{Ragan-Kelley2012}, to program \mempool{} with high-level abstractions (\cref{sec:programming_model});
  \item An evaluation of \mempool{}'s performance and architectural bottlenecks with a wide set of \gls{dsp} kernels and full-blown applications from a wide range of domains (\cref{sec:evaluation});
  \item The open-source release of \mempool{}, including open simulator support and the full software infrastructure\footnote{Available on \emph{\url{https://github.com/pulp-platform/mempool}}.}.
\end{itemize}

In GlobalFoundries 22FDX technology, \mempool{} runs at \SI{600}{\mega\hertz}, approximately 60 gate delays, in typical operating conditions (TT/\SI{0.80}{\volt}\kern-.1em/\SI{25}{\celsius}). \mempool{}'s L1 interconnect keeps the average latency at fewer than six cycles, even for a heavy interconnect load of \SI{0.35}{\request\per\core\per\cycle}. Our hybrid addressing scheme helps to reduce the average latency and power consumption further. We demonstrate energy savings up to \SI{28}{\percent} through various optimizations to the instruction cache with respect to our previous work~\cite{Cavalcante2021}. Furthermore, we illustrate how to efficiently refill all instruction caches through an \gls{axi} tree interconnect enhanced with software-controlled caches. \mempool{}'s \gls{axi} interconnect offers a bandwidth of \SI{256}{\byte\per\cycle}, which is fully utilized by our distributed \gls{dma}. We developed compiler support in GCC and LLVM to hide architectural latencies to the L1 \gls{spm} through instruction scheduling. Thanks to our OpenMP runtime and Halide framework, the implementation of complex applications is straightforward. \mempool{} achieves a performance up to \SI{229}{\giga\ops} and an energy efficiency up to \SI{180}{\giga\ops\per\watt}. On average, the memory system only causes stalls \SI{4}{\percent} of the execution time, and we achieve \SI{0.96}{\ipc} on \mempool{}'s \glspl{pe}.

The remainder of this paper is structured as follows. It starts with an overview of \mempool{}'s general architecture and processor design in \cref{sec:architecture_overview}. Next, we explain individual components and justify their configuration: \cref{sec:l1_memory_interconnect} describes \mempool{}'s L1 data interconnect, and the associated hybrid addressing scheme; \Cref{sec:instruction_cache} describes \mempool{}'s instruction cache and its optimization steps; finally, we explain the system interconnect in \cref{sec:system_interconnect}, including the instruction cache hierarchy and the \gls{dma} engine. We present \mempool{}'s physical implementation in \cref{sec:implementation} before defining its programming model in \cref{sec:programming_model} and evaluate its performance in \cref{sec:evaluation}. Finally, related work and conclusion are discussed in Sections \ref{sec:related_work} and \ref{sec:conclusion}.

%%%%%%%%%%%%%%%%%%%%%%%%%%%%%%%%%%%%%%%%%%%%%%%%%%%%%%%%%%%%%%%%%%%%%%%%%%%%%%%
%
% Architecture
%
%%%%%%%%%%%%%%%%%%%%%%%%%%%%%%%%%%%%%%%%%%%%%%%%%%%%%%%%%%%%%%%%%%%%%%%%%%%%%%%

\begin{figure*}[th]
  \centering
  \includegraphics[width=\linewidth]{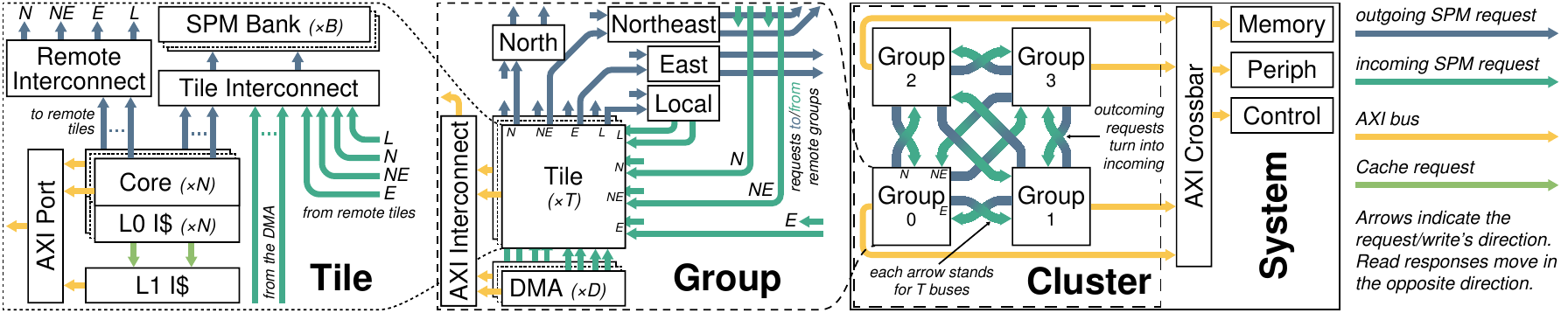}
  \caption{Bottom-up overview of \mempool{}'s architecture highlighting its hierarchy and interconnects. From left to right, it starts with the \emph{tile}, which holds $N$ cores with private L0 and a shared L1 instruction cache, $B$ \gls{spm} banks, and remote connections. The \emph{group} features $T$ such tiles and a \emph{local} L1 interconnect to connect tiles within the group, as well as a \emph{north}, \emph{northeast}, and \emph{east} L1 interconnect connecting to other groups. It also holds a hierarchical \gls{axi} interconnect and \glspl{dma}. The cluster contains four groups and connects to the system's memory and peripherals through an \gls{axi} crossbar. This paper focuses mainly on a large \mempool{} configuration with $T=16$ tiles per group with $N=4$ cores and $B=16$ banks each.}\label{fig:arch_overview}
\end{figure*}

\section{MemPool Architecture Overview}%
\label{sec:architecture_overview}

\mempool{} is a flexible and parametric manycore architecture with hundreds of individually programmable cores sharing a low-latency L1 \acrfull{spm}. \changed{Note, the L1 \gls{spm} is not a hardware cache but software-managed, physically addressed memory.} Its architecture is shown in \cref{fig:arch_overview} and detailed in a bottom-up fashion in the subsequent sections.

We use the Snitch core~\cite{Zaruba2020} as \gls{pe}, a small single-stage 32-bit processor implementing \riscv{}'s RV32IMAXpulpimg \gls{isa}. A distinctive feature of Snitch is the capability to handle multiple outstanding instructions: this is key to tolerating the latency of load and stores operation in \mempool{}. In \cref{sec:processor}, we describe our modifications to Snitch and its key features in detail.

The first building block of \mempool{}'s hierarchy is the \emph{tile}. It combines a few cores, the first levels of the instruction cache, and a subset of the shared L1 \gls{spm} banks connected with a fully connected crossbar to the cores. Each tile also has a parameterizable number of \emph{remote} ports, which allow the cores to make requests into remote tiles' \gls{spm}. Analogously, each tile has incoming request ports to serve the memory requests from remote tiles. Those ports connect directly to the tile's fully connected crossbar. Incoming and outgoing request ports can be pipelined to reduce the critical path at the cost of latency. \Cref{sec:l1_memory_interconnect} explains the \gls{pe} to L1 \gls{spm} interconnect in detail. Every tile has its own instruction cache consisting of a shared L1 cache and small, tightly-coupled L0 caches that are private to each core, as explained in \Cref{sec:instruction_cache}. The tile also features an \gls{axi} port connecting the cores and the cache to a system bus and memory in the upper hierarchy.

Multiple tiles are combined and connected in a \emph{group}. Each tile's remote port connects to one of several interconnects, which routes the requests to the target group and its tile's \gls{spm}. The group also contains a tree-like \gls{axi} interconnect and instruction cache hierarchy to connect to the system bus. The \gls{axi} connection to the system, including its cache hierarchy and \gls{dma}, are further described in \cref{sec:system_interconnect}.

Several groups compose a \emph{cluster} with point-to-point \gls{spm} interconnect connections. The cluster also forwards the groups' \gls{axi} interconnect to the \gls{soc}. On a system level, \mempool{} connects to a long-latency (tens to hundreds of cycles) L2 memory, e.g., a large on-chip memory or an off-chip \gls{dram}.

%%%%%%%%%%%%%%%%%%%%%%%%%%%%%%%%%%%%%%%%%%%%%%%%%%%%%%%%%%%%%%%%%%%%%%%%%%%%%%%
%
% Compute unit specialization
%
%%%%%%%%%%%%%%%%%%%%%%%%%%%%%%%%%%%%%%%%%%%%%%%%%%%%%%%%%%%%%%%%%%%%%%%%%%%%%%%

\subsection{Core}\label{sec:processor}

To scale \mempool{}, its cores must have a small footprint, tolerate and hide the L1 interconnect latency, and be extensible. Snitch, a single-stage 32-bit \riscv{} core, meets these requirements~\cite{Zaruba2020}. While Snitch is a single-issue core, it features a scoreboard supporting multiple outstanding instructions, which is crucial for two reasons. First, it allows issuing multiple outstanding load or store requests without blocking the pipeline if there are no \gls{raw} dependencies. Second, Snitch features an accelerator port to add complex, pipelined functional units such as a \gls{mac} unit. Snitch can offload suitable instructions to those functional units and continue its operation.

We extend Snitch for \mempool{} by adding the capability to retire load responses out-of-order or reorder them due to \mempool{}'s \gls{numa} interconnect not providing response ordering. By supporting eight outstanding transactions, Snitch can completely hide the L1 interconnect's five cycles of latency through instruction scheduling with headroom for further load-induced latency.

\changed{\mempool{}'s \riscv{} cores stipulate the \gls{rvwmo} memory model\footnote{See \url{https://riscv.org/technical/specifications/} for more details.}. While cores observe their own stores in program order, the stores of other cores can be observed reordered. The \emph{fence} instructions allow the programmers to enforce a memory order between cores. Furthermore, \mempool{} is designed as an accelerator cluster with Snitch as simple compute cores. Hence, we do not consider running an operating system nor having external interrupts or context switches. Finally, we do not support virtual memory, and the L1 interconnect will never return error responses. This removes the need for exception handling and allows Snitch to retire instructions out-of-order without complex additional hardware. Those features could be enabled through a reorder buffer inserted into Snitch's \gls{lsu} to retire loads in order.}

\changed{We focus on integer processing and configure Snitch to support the RV32IMAXpulpimg set.} The \emph{Xpulpimg} instruction set~\cite{Mazzola2021} is supported by a pipelined \gls{ipu} connected to Snitch's accelerator port. Its instructions like \gls{mac} and load-post-increment drastically reduce the instruction count for \gls{dsp} kernels and result in a significant speedup. We extend Snitch's register file to have three read ports to supply all the operands necessary for \gls{mac} operations and two write ports to reduce contention when retiring instructions with different latencies.

\subsection{\mempool{} Configuration}

While the \mempool{} design is highly parametric, for clarity, we mainly focus on the largest and most challenging configuration from the physical implementation viewpoint. Specifically, this maximum configuration contains \num{256} cores distributed across four groups. Each group consists of 16 tiles containing four cores each. The whole cluster features \num{1024} \SI{1}{\kibi\byte} \gls{sram} banks each, resulting in \SI{1}{\mebi\byte} of \gls{spm} and a banking factor of 4. Each tile has 32 instructions per core's private L0 cache and \SI{2}{\kibi\byte} of two-way set-associative, shared L1 instruction cache. We configure the \gls{axi} data width to be \SIadj{512}{\bit} wide and parameterize the hierarchical \gls{axi} interconnect to result in one \gls{axi} master port per group. Each group contains one \SI{8}{\kibi\byte} \gls{ro} cache. This large \mempool{} configuration features significantly more \glspl{pe} and larger L1 \gls{spm} than most common L1-coupled clusters, which range from 8 \glspl{pe} and \SI{128}{\kibi\byte} of L1 \gls{spm}~\cite{Zaruba2021} to 64 \glspl{pe}~\cite{Ginosar2016}.

%%%%%%%%%%%%%%%%%%%%%%%%%%%%%%%%%%%%%%%%%%%%%%%%%%%%%%%%%%%%%%%%%%%%%%%%%%%%%%%
%
% MemPool's interconnect
%
%%%%%%%%%%%%%%%%%%%%%%%%%%%%%%%%%%%%%%%%%%%%%%%%%%%%%%%%%%%%%%%%%%%%%%%%%%%%%%%

\section{L1 Memory Interconnect}%
\label{sec:l1_memory_interconnect}

This section elaborates on how to connect cores to L1 \gls{spm}. We explore three interconnect topologies, enhance them with a hybrid addressing scheme, and compare their performance.

Designing a low-latency L1 interconnect that links 256 \glspl{pe} (initiators) with 1024 memory banks (targets) is a major challenge. High memory throughput, low access latency, and a shared view of memory are required while minimizing the area overhead and ensuring that the interconnect is physically implementable. 1024 \SI{1}{\kibi\byte} \gls{sram} banks form the \SI{1}{\mebi\byte} L1. The banking factor of \num{4} helps to reduce the probability of banking conflicts without requiring the programmer to schedule each core's bank accesses explicitly.

\subsection{Interconnect Topologies}\label{subsec:interconnect_topologies}

The straightforward implementation of the L1 memory interconnect is a fully connected \by{256}{1024} crossbar. It provides the highest possible throughput thanks to dedicated paths between each master-slave pair eliminating internal routing conflicts. Unfortunately, a crossbar has a quadratic area cost~\cite{Dally2004}, obviously making it prohibitively expensive at \mempool{}'s scale. Instead, we implement this interconnect hierarchically, with \emph{tiles} containing cores and \gls{sram} banks connected at higher hierarchy levels. \cref{fig:arch_overview} shows the tile architecture. It contains four cores and \num{16} L1 \gls{sram} banks of \SI{1}{\kibi\byte} each, accessible in a single cycle by the tile's cores.

\begin{figure}[tb]
  \centering
  \includegraphics[]{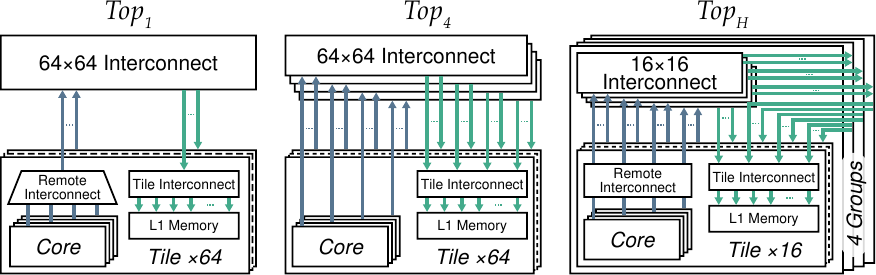}
  \caption{Evaluated interconnect topologies highlighting the connection between cores and L1 \gls{spm} in remote tiles.}
  \label{fig:arch_topologies}
\end{figure}

The \mempool{} cluster contains \num{64} tiles, totaling \num{256} cores and \SI{1}{\mebi\byte} of L1 \gls{spm}. In \cite{Cavalcante2021}, we proposed and evaluated the three topologies shown in \cref{fig:arch_topologies}, balancing the design's physical feasibility, access latency, and throughput.

\topology{1} has a single outgoing and incoming remote port per tile. The tiles are connected with a \by{64}{64} radix-4 butterfly network, with a single pipeline stage midway through its $\log_4 (64) = 3~\text{layers}$, resulting in a latency of \num{5}~cycles for remote accesses. Inside the tile, the \by{5}{16} fully connected \emph{tile interconnect} connects the local cores and incoming port to the \num{16} banks. The single remote port becomes the bottleneck since four cores compete for access. \topology{4} tackles this bottleneck with four remote request ports, one per core. The tiles are connected by four independent \by{64}{64} radix-4 butterfly networks with a core-to-memory latency of \num{5}~cycles.

\topology{H} also has four outgoing and incoming remote request ports per tile. However, to make this topology feasible, we exploit the physical proximity of the different tiles and add another hierarchy level to \mempool{}, the \emph{group}. It consists of 16 tiles, a fourth of the \mempool{} cluster. Cores can access remote data in the same group with a latency of \num{3}~cycles through a \by{16}{16} fully connected crossbar internal to the group. Moreover, each group pair is connected through an independent \by{16}{16} fully connected crossbar, with a latency of \num{5}~cycles. Therefore, each group has four \by{16}{16} fully connected crossbars, one to connect the tiles within the same group (\emph{local} interconnect) and three others to connect to remote groups (\emph{north}, \emph{northeast}, and \emph{east} interconnects).

%%%%%%%%%%%%%%%%%%%%%%%%%%%%%%%%%%%%%%%%%%%%%%%%%%%%%%%%%%%%%%%%%%%%%%%%%%%%%%%
%
% Hybrid addressing scheme
%
%%%%%%%%%%%%%%%%%%%%%%%%%%%%%%%%%%%%%%%%%%%%%%%%%%%%%%%%%%%%%%%%%%%%%%%%%%%%%%%

\subsection{Hybrid Addressing Scheme}\label{sec:hybrid_addressing_scheme}

The L1 interconnect topologies are the first step towards a shared-memory manycore system. However, the opportunity to access remote tiles' memory comes with trade-offs. For all topologies, remote requests have lower throughput, higher latency, and consume more power. We propose a \emph{hybrid addressing scheme} to mitigate those effects by keeping as many memory requests as possible within a tile.

\begin{figure}[tb]
  \centering
  \includegraphics[width=\linewidth]{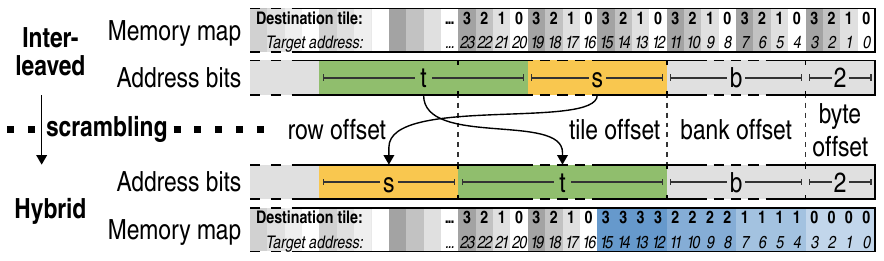}
  \caption{Address scrambling transforming a fully interleaved memory map (top) to a hybrid one (bottom). The outer bars visualize the memory map with interleaved regions in gray and sequential regions in blue. For simplicity, we assume only four different tiles represented by shades.}
  \label{fig:scrambling_logic}
\end{figure}

\mempool{} has a word-level interleaved memory mapping across all memory banks to minimize banking conflicts. However, this implies that most memory requests target remote tiles. Ideally, most requests remain in the local tile to minimize latency and power consumption. With the scrambling logic of \cref{fig:scrambling_logic}, we transform an interleaved memory map into a \emph{hybrid} one by creating \emph{sequential regions} (in blue) in which contiguous addresses target a single tile.

In an interleaved memory addressing scheme, the addresses are interpreted as follows. The first two bits are the byte offset, after which $b$ bits identify one of each tile's $2^{b}$ banks. The next $t$ bits distinguish between the $2^{t}$ tiles. The remaining bits define the row offset within the bank. In a continuous addressing scheme, lower address bits represent the bank's row offset, and the upper bits select a bank. We dedicate $2^{s}$ rows of each tile's banks, or \SI[parse-numbers=false]{2^{s+b+2}}{\byte} in each tile, to the sequential memory region. To keep accesses within the same tile interleaved, we leave the byte and bank offsets untouched. The next $s$ bits represented part of the tile offset but should now define the banks' next row within the same tile. Hence, we shift them $t$ bits to the left to the start of the row offset. Incrementing these address bits will consequently traverse the rows of banks within one tile, while the tile offset should stay constant. Therefore, we fill the shifted bits with the $t$ bits they replaced. This permutation creates $2^{t}$ sequential regions, one for each tile. In total, we dedicate the first \SI[parse-numbers=false]{2^{t+s+b+2}}{\byte} to sequential regions. We leave the subsequent bytes interleaved by only scrambling addresses inside the sequential memory region. The scrambling logic can be efficiently implemented in hardware with a wire crossing and a multiplexer.

\changed{The hybrid addressing scheme's key benefit is to allow the programmer to store private data, such as the stack, locally in the \gls{pe}'s tile. This is done automatically by the runtime. Furthermore, we provide specific allocator calls to allocate dynamic memory in local regions. This greatly reduces the number of transactions between tiles, making better use of the tiles' local high-throughput L1 crossbar. In contrast to aliased, fully private memories, we do not complicate programmability. We give all \glspl{pe} the same memory view and keep the L1 memory region contiguous. Programs that heavily use the stack or work mainly on local data immensely benefit from the sequential regions.}

\subsection{Evaluation}

\subsubsection{L1 Memory Interconnect}

We analyze the physical feasibility of the presented topologies presented. We find that both \topology{1} and \topology{H} are feasible, while \topology{4} is not, because of the large routing requirements of its four \by{64}{64} global interconnects~\cite{Cavalcante2021}. Traffic generators, which generate new requests following a Poisson process of rate $\lambda$, replace the cores to analyze the topologies' latency and throughput as a function of the injected load $\lambda$ (measured in requests per core per cycle). The requests have a random, uniformly distributed destination bank.

\begin{figure}[bt]
  \centering
  \begin{minipage}[ht]{0.5\linewidth}
    \resizebox{\linewidth}{!}{
    \begin{tikzpicture}
      \begin{axis}[
        xlabel = {Injected load (\si{\request\per\core\per\cycle})},
        xmin = 0,
        xmax = 0.5,
        xtick distance = 0.1,
        ylabel = {Throughput (\si{\request\per\core\per\cycle})},
        ymin = 0,
        ymax = 0.5,
        ytick distance = 0.1,
        height = 6cm,
        grid   = major,
        legend pos = north west,
        legend style = {font=\footnotesize}]

        \addplot [color3, ultra thick] table [x expr = \thisrowno{0}*0.001] {fig/interco/throughput_toph};
        \addlegendentry{\topology{H}};
        \addplot [color2, ultra thick] table [x expr = \thisrowno{0}*0.001] {fig/interco/throughput_top4};
        \addlegendentry{\topology{4}};
        \addplot [color1, ultra thick] table [x expr = \thisrowno{0}*0.001] {fig/interco/throughput_top1};
        \addlegendentry{\topology{1}};
      \end{axis}
    \end{tikzpicture}}
    % \subcaption{Throughput.}
    % \label{fig:thru}
  \end{minipage}\hfill%
  \begin{minipage}[ht]{0.5\linewidth}
    \resizebox{\linewidth}{!}{
    \begin{tikzpicture}
      \begin{axis}[
        xlabel = {Injected load (\si{\request\per\core\per\cycle})},
        xmin = 0,
        xmax = 0.5,
        xtick distance = 0.1,
        ylabel = {Average latency (\si{\cycle})},
        ymin = 0,
        ymax = 20,
        restrict y to domain=*0:200,
        ytick distance = 4,
        height = 6cm,
        grid   = major]

        \addplot [color1, ultra thick] table [x expr = \thisrowno{0}*0.001] {fig/interco/latency_top1};
        \addplot [color2, ultra thick] table [x expr = \thisrowno{0}*0.001] {fig/interco/latency_top4};
        \addplot [color3, ultra thick] table [x expr = \thisrowno{0}*0.001] {fig/interco/latency_toph};
      \end{axis}
    \end{tikzpicture}}
    % \subcaption{Average latency.}
    % \label{fig:latency}
  \end{minipage}
  \caption{Network analysis of the three proposed topologies, in terms of throughput and average round-trip latency, as a function of the load.}
  \label{fig:analysis}
\end{figure}
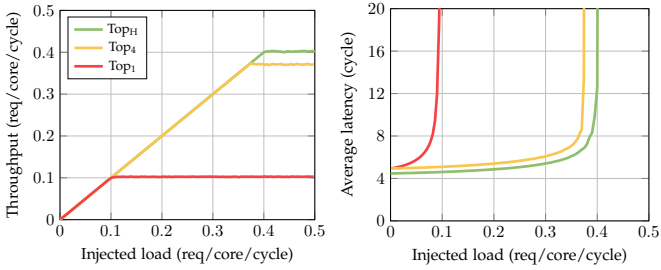

\cref{fig:analysis} shows the throughput and latency of different topologies. At an injected load of \SI{0.10}{\request\per\core\per\cycle}, \topology{1} becomes congested, while \topology{4} and \topology{H} support almost four times that load, about \SI{0.37}{\request\per\core\per\cycle} and \SI{0.4}{\request\per\core\per\cycle}, respectively. \topology{H}'s throughput is slightly higher than \topology{4}'s due to its smaller diameter.
The explosion of the average latency highlights the point where the topologies become congested. \topology{H}'s average latency only reaches \num{6} cycles at a network load of \SI{0.35}{\request\per\core\per\cycle}. Due to \topology{H}'s three-cycle latency to a local group, it achieves a smaller average latency than \topology{4}. Since \topology{H} clearly outperforms \topology{1} as well as \topology{4}, which is physically infeasible, we implement \topology{H} as \mempool{}'s L1 interconnect.

\subsubsection{Hybrid Addressing Scheme}

To illustrate the benefit of the hybrid addressing scheme, we analyze \topology{H} taking the hybrid addressing scheme into account. The traffic generator creates uniformly-distributed requests to the local tile's sequential region with probability $p_{\text{local}}$, and outside of this region with probability $1-p_{\text{local}}$.

\begin{figure}[bt]
  \centering
  \begin{minipage}[ht]{0.5\linewidth}
    \resizebox{\linewidth}{!}{
    \begin{tikzpicture}%[/tikz/font=\footnotesize]
      \begin{axis}[
        xlabel = {Injected load (\si{\request\per\core\per\cycle})},
        xmin = 0,
        xmax = 1,
        xtick distance = 0.2,
        ylabel style={align=left},
        ylabel = {Throughput (\si{\request\per\core\per\cycle})},
        ymin = 0,
        ymax = 1,
        restrict y to domain=*0:1,
        ytick distance = 0.2,
        height = 6cm,
        grid   = major]

        \addplot [color1, ultra thick] table [x expr = \thisrowno{0}*0.001] {fig/scrambling/throughput_0};
        \addplot [color2, ultra thick] table [x expr = \thisrowno{0}*0.001] {fig/scrambling/throughput_250};
        \addplot [color3, ultra thick] table [x expr = \thisrowno{0}*0.001] {fig/scrambling/throughput_500};
        \addplot [color4, ultra thick] table [x expr = \thisrowno{0}*0.001] {fig/scrambling/throughput_750};
        \addplot [color5, ultra thick] table [x expr = \thisrowno{0}*0.001] {fig/scrambling/throughput_1000};
      \end{axis}
    \end{tikzpicture}}
    % \subcaption{Throughput.}
    % \label{fig:thru_scr}
  \end{minipage}\hfill%
  \begin{minipage}[ht]{0.5\linewidth}
    \resizebox{\linewidth}{!}{
    \begin{tikzpicture}%[/tikz/font=\footnotesize]
      \begin{axis}[
        xlabel = {Injected load (\si{\request\per\core\per\cycle})},
        xmin = 0,
        xmax = 1,
        xtick distance = 0.2,
        ylabel = {Average latency (\si{\cycle})},
        ymin = 0,
        ymax = 20,
        restrict y to domain=*0:300,
        ytick distance = 4,
        height = 6cm,
        grid   = major,
        legend pos = north west,
        legend style = {font=\footnotesize}]]

        \addplot [color1, ultra thick] table [x expr = \thisrowno{0}*0.001] {fig/scrambling/latency_0};
        \addlegendentry{\SI{0}{\percent}}
        \addplot [color2, ultra thick] table [x expr = \thisrowno{0}*0.001] {fig/scrambling/latency_250};
        \addlegendentry{\SI{25}{\percent}}
        \addplot [color3, ultra thick] table [x expr = \thisrowno{0}*0.001] {fig/scrambling/latency_500};
        \addlegendentry{\SI{50}{\percent}}
        \addplot [color4, ultra thick] table [x expr = \thisrowno{0}*0.001] {fig/scrambling/latency_750};
        \addlegendentry{\SI{75}{\percent}}
        \addplot [color5, ultra thick] table [x expr = \thisrowno{0}*0.001] {fig/scrambling/latency_1000};
        \addlegendentry{\SI{100}{\percent}}
      \end{axis}
    \end{tikzpicture}}
    % \subcaption{Average latency.}
    % \label{fig:latency_scr}
  \end{minipage}
  \caption{Network analysis of \topology{H} with our hybrid addressing scheme for different probabilities of requesting data in the sequential region $p_{\text{local}}$.}
  \label{fig:analysis_scramble}
\end{figure}
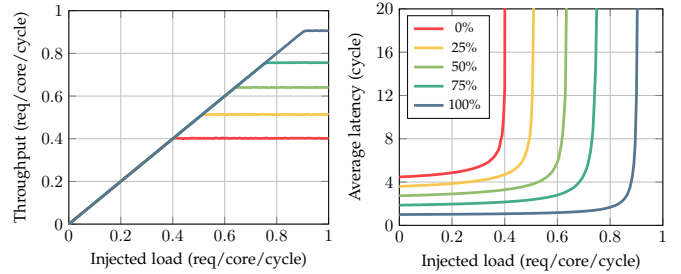

\cref{fig:analysis_scramble} shows the throughput and latency of \topology{H} for different $p_{\text{local}}$. It shows a clear trend of an increased throughput for a larger $p_{\text{local}}$. The scrambling logic can vastly improve the system's throughput by preventing congestion in the global interconnect, besides lowering the overall average access latency. An application making \SI{25}{\percent} of its accesses to the stack can gain up to \SI{27}{\percent} in performance with the hybrid addressing scheme without changing the code. Due to the clear benefit of using the hybrid addressing scheme, we always enable it in \mempool{}.

%%%%%%%%%%%%%%%%%%%%%%%%%%%%%%%%%%%%%%%%%%%%%%%%%%%%%%%%%%%%%%%%%%%%%%%%%%%%%%%
%
% Instruction path
%
%%%%%%%%%%%%%%%%%%%%%%%%%%%%%%%%%%%%%%%%%%%%%%%%%%%%%%%%%%%%%%%%%%%%%%%%%%%%%%%

\section{Instruction Cache}\label{sec:instruction_cache}

With the cores connected to the \gls{spm}, we must address their instruction path. This section describes the implementation and optimization of the tile's instruction cache.

\subsection{Implementation}

% L0 cache
Since Snitch is a single-stage processor without a pipelined instruction fetch stage, the instruction cache lookup's combinational path is stacked on top of the path crossing the entire processor. Therefore, each processor has a minimal, private, fully associative, \gls{scm}-based L0 cache to minimize the impact on the critical path. Its line width and count are configurable. To avoid the miss penalty of such a small cache, it implements prefetching, which gets the next line by scanning the current cache line for backward branches (representing loops) or predictable jumps.

% L1 cache
A shared L1 cache per tile feeds the L0 caches. It implements a configurable set-associative cache lookup where the data and tag banks can be implemented as \gls{sram} macros or \gls{scm}. Its refill logic coalesces outstanding refill requests and responds to all L0 caches in parallel to reduce latency. The refill port is connected to the tile's wide \gls{axi} port.

We optimize and evaluate the following cache configurations and measure the impact on speed and power with cycle-accurate post-layout power simulations:

\begin{description}
\sisetup{detect-all = true}

\item[Baseline:] \emph{(\SI{149}{\kilo\ge})}
The cache architecture used in the preliminary \mempool{} version~\cite{Cavalcante2021}. The L0 cache is register-based, and each core has four \SIadj{128}{\bit} lines. The L1 cache implements a \SI{2}{\kibi\byte}, 4-way set-associative, parallel lookup where the data, as well as the tag banks, are \gls{sram}-based. For a lookup, we read the complete set (all the tag and data banks) simultaneously and then select the correct cache line from the hit calculation.

\item[2-Way:] \emph{(\SI{163}{\kilo\ge})}
The cache line width is doubled to \SI{256}{\bit}. The key idea behind this change is to alleviate pressure on the L1 interface. Previously, each of the four L0 caches had to request a new cache line every four instructions executed, fully utilizing the L1 interface unless the code completely fits into the L0 cache. With eight instructions per line, each L0 cache only requires a new cache line every eight instructions, leaving room for mispredicted branches and jumps. We reduce the associativity to two to compensate for the wider \gls{sram} banks in the L1 cache. In combination, this keeps the L1 cache size constant while doubling the L0 cache, which allows bigger kernels to fit entirely in the L0 cache, reducing instruction stalls and L1 power consumption.

\item[L1-Tag Latch:] \emph{(\SI{161}{\kilo\ge})}
This configuration replaces the \gls{sram}-based tag banks of the L1 cache with latch-based \glspl{scm} to reduce power and area (due to its small size).

\item[L1-All Latch:] \emph{(\SI{217}{\kilo\ge})}
Going a step further, we also replace the L1 cache's data banks with latch-based \glspl{scm}. This drastically increases area as the data banks are big enough to benefit from being implemented in \gls{sram}. We discard this solution due to the significant area overhead, which conflicts with our physical constraints.

\item[L1-Tag+L0 Latch:] \emph{(\SI{153}{\kilo\ge})}
Rather than making the L1 data latch-based, we replace the registers in the L0 cache with latches for area and power reduction.

\item[Serial L1:] \emph{(\SI{123}{\kilo\ge})}
The last architecture implements a serial L1 lookup where first the \gls{scm}-based tags are checked in parallel, and then the data from the way that hit is read. This allows merging the \gls{sram} banks of both data ways into a single bank, which is more area-efficient than two banks that can be read in parallel and conserves energy by minimizing the number of \gls{sram} accesses. This serial lookup is fully pipelined, keeping the throughput constant, but inherently adds an extra cycle of lookup latency. However, prefetching the L0 cache can hide this latency during regular operation.

\end{description}

\subsection{Evaluation}\label{sec:instruction_path_optimization}

To evaluate the cache architectures, we analyze their power consumption using the methodology described in \cref{sec:methodology}. We use two different workloads taken from our benchmarks described in \cref{subsec:microarch_benchmarking} to give an exemplary quantitative assessment of the different cache organization options: 1) a \smallkernel{} kernel that fits into the optimized L0 cache and 2) a \bigkernel{} kernel that never fits into the L0 cache. \changed{The \smallkernel{} kernel has a \SI{2}{\percent} speedup from doubling the L0 cache size in the \emph{2-Way} optimization. Changing the memory type to \gls{scm} does not impact functionality or performance. Since this kernel fits into the L0 cache, there is no further performance impact by the final L1 cache optimization. The \bigkernel{} kernel gains \SI{5}{\percent} of performance with the \emph{2-Way} optimization. The  \emph{Serial L1} configuration's additional latency reduces this gain to \SI{2}{\percent} compared to the \emph{Baseline}. Performance changes are small because the L0 cache's prefetching hides most misses in all configurations. Both kernels achieve a speedup of \SI{2}{\percent} with the final architecture compared to the \emph{Baseline}.}

\begin{figure}[bth]
  \centering
  \includegraphics[]{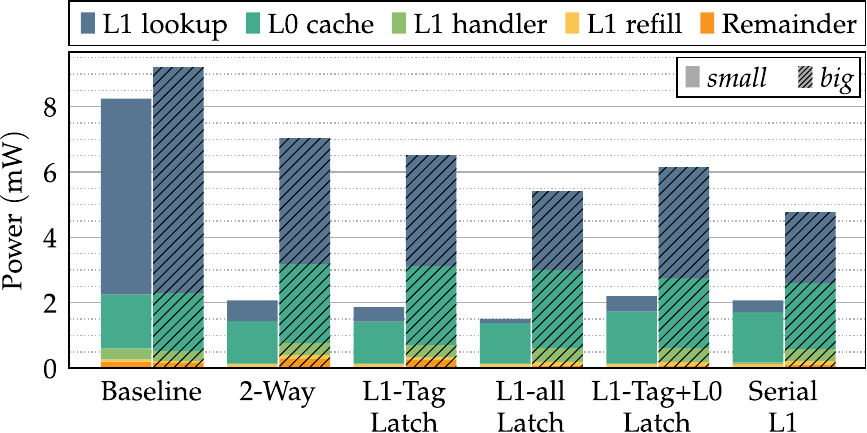}
  \caption{Breakdown of the instruction cache's power consumption at various optimization steps for a \smallkernel{} kernel that fits into the L0 cache and a \bigkernel{} one that does not.}
  \label{fig:cache_power}
\end{figure}

\begin{figure}[bth]
  \centering
  \includegraphics[]{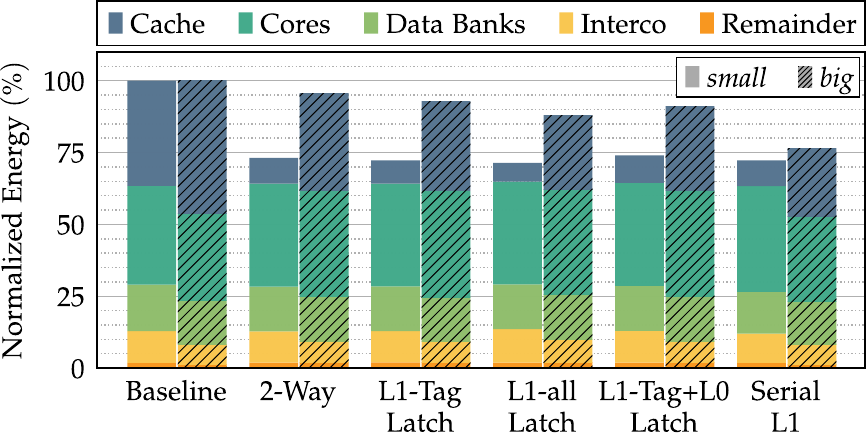}
  \caption{Breakdown of the tile's normalized energy consumption at various optimization steps for a \smallkernel{} kernel that fits into the L0 cache and a \bigkernel{} one that does not.}
  \label{fig:tile_energy}
\end{figure}

\Cref{fig:cache_power} shows the cache architectures' power consumption. In the baseline version, the power consumption is dominated by the \gls{sram} banks. Increasing the cache size to make the kernel fit into L0 diminishes the L1 lookup power for the \smallkernel{} kernel, and in combination with the later optimizations, it saves \SI{6.2}{\milli\watt} (\SI{-75}{\percent}). The \bigkernel{} kernel, which continuously has to be refilled from L1, stresses the benefits of the later optimizations. The L1 cache's power consumption decreases by \SI{45}{\percent} when reducing the associativity and thereby halving the \gls{sram} reads. Moving to latch-based \glspl{scm} reduces power further. Finally, the serial lookup L1 cache architecture reduces its consumption by \SI{36}{\percent}. Overall, the optimized cache saves \SI{4.5}{\milli\watt} (\SI{-48}{\percent}) per tile.

The tile's energy consumption is shown in \cref{fig:tile_energy}. The slight increase in the cores' energy consumption suggests the performance increase thanks to the larger L0 cache. Energy efficiency gains of \SI{28}{\percent} and \SI{24}{\percent} for \smallkernel{} and \bigkernel{} kernel, respectively, accentuate the key role of an efficient cache subsystem in a fully programmable manycore system.

To summarize, we double the L0 cache size from \num{16} to \num{32} instructions by going to a \SIadj{256}{\bit}-wide cache line; we move from 4-way to 2-way set associativity and go from a parallel to a serial lookup architecture in the L1 cache; we change the L0 cache and the L1 cache's tag to be latch-based while keeping the L1 data \gls{sram}-based. Overall, we reduce the cache area by \SI{17}{\percent} and improve the performance of our benchmarks by \SI{2}{\percent}. We save \SI{5.9}{\milli\watt} (\SI{-75}{\percent}) in one tile when the kernel fits into L0; otherwise, we save \SI{4.4}{\milli\watt} (\SI{-48}{\percent}). For the full \mempool{} with 64 tiles, this amounts to \SI{378}{\milli\watt} and \SI{282}{\milli\watt}, respectively.

\section{System Interconnect and DMA engine}\label{sec:system_interconnect}

With the tile's cache being optimized and the cores having access to a shared L1 data memory, we need to make sure the caches can refill and that we can transfer data to and from the system memory. Giving each of the 64 tiles a private high-bandwidth connection to the system is exceptionally challenging in an already routing-dominated design. Note, the preliminary version of \mempool{}~\cite{Cavalcante2021} did not consider such an interconnect and instead assumed an ideal connection between the tiles and system memory. This section tackles the challenge of connecting all 64 tiles' \gls{axi} port to the system bus with minimal area and routing overhead. Specifically, we need to extend the instruction cache hierarchy to sustain the cores' instruction stream from L2 or external memory, and we need to design a special \gls{dma} that can transfer data between \mempool{}'s distributed L1 \gls{spm} and L2 or external memory with minimal overhead and without congesting the cores' interconnects.

\subsection{Hierarchical AXI interconnect}

The instruction and data path are latency tolerant and take advantage of spatial locality by operating on contiguous memory chunks. Therefore, we use the open and standard \gls{axi} protocol to connect the tiles to the system, as it is burst-based, latency-tolerant, multi-initiator/target with decoupled read and write channels.

\begin{figure}[b]
  \centering
  \includegraphics[]{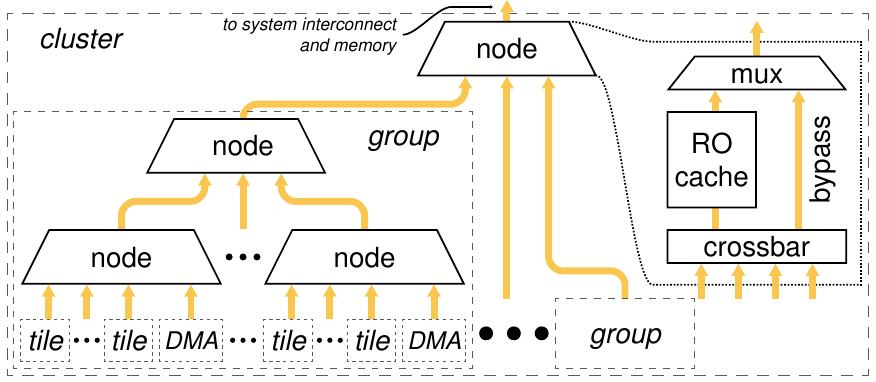}
  \caption{Hierarchical \gls{axi} interconnect with optional read-only caches.}
  \label{fig:hierarchiacal_axi}
\end{figure}

Each tile has an \gls{axi} master port shared between all cores and the instruction cache refill, giving them access to higher-level/main memory, peripherals, and control registers. However, routing a private \gls{axi} bus per tile to the cluster boundary is not physically feasible. For this reason, we designed the \gls{axi} interconnect with a hierarchical approach as shown in \cref{fig:hierarchiacal_axi}. Each tile and \gls{dma} is a leaf node in a configurable \gls{axi} tree. At each level in the interconnect, neighboring child nodes merge into a single \gls{axi} bus. At the top level, we keep a configurable number of master ports to allow for high bandwidth, effectively building multiple \gls{axi} trees with a subset of the tiles and \glspl{dma} each.

\subsection{Read-only cache}

Combining multiple \gls{axi} masters not only reduces routing but also allows for caches to coalesce requests at each level to reduce the bandwidth requirements. To this end, we implement a specialized \gls{ro} cache and optionally instantiate it at each node. We leave out write-support to the cache to keep it small and energy-efficient by avoiding coherency problems. The cache is software-managed, giving the programmer full control over cached regions and cache flushes. This enables caching the binary and \gls{ro} data during booting and by the programmer.

The \gls{ro} cache consists of four stages. In the \emph{\gls{axi} to cache} stage, \gls{axi} bursts are broken down into individual cache requests, and their metadata is stored to reconstruct the correct \gls{axi} responses later. The cache requests traverse the \emph{lookup} stage, whose output is processed by the \emph{handler}. In case of a hit, data is returned to the \emph{\gls{axi} to cache} stage to issue \gls{axi} responses. For a miss, the handler checks if a corresponding refill is in flight or issues a new one. The cache supports multiple outstanding requests. However, the \gls{axi} protocol imposes one constraint: requests of the same ID must return in order. Since hits can quickly overtake misses, they must be stalled if the same master issued a pending miss earlier.

The \gls{ro} cache's depth and width can be configured and are design parameters that define the bandwidth and the amount of coalescing the cache provides. The cache is designed to be fully pipelined to allow for maximum throughput. We will primarily use the \gls{ro} cache for instructions since we rarely want to cache \gls{dma} transfers or move big chunks of \gls{ro} data with the processors. Hence, we tune it mainly for the instruction path while also dimensioning the hierarchical \gls{axi} interconnect to serve the \gls{dma} well.

%%%%%%%%%%%%%%%%%%%%%%%%%%%%%%%%%%%%%%%%%%%%%%%%%%%%%%%%%%%%%%%%%%%%%%%%%%%%%%%
%
% DMA
%
%%%%%%%%%%%%%%%%%%%%%%%%%%%%%%%%%%%%%%%%%%%%%%%%%%%%%%%%%%%%%%%%%%%%%%%%%%%%%%%
\subsection{DMA engine}\label{sec:dma_engine}

The final piece of \mempool{}'s architecture is a \gls{dma} engine to move data to and from L1 memory efficiently, utilizing the \gls{axi} interconnect and having access to all 1024 \gls{spm} banks.

Due to \mempool{}'s distributed L1 \gls{spm} and large scale, a classic \gls{dma} engine is not feasible. Since \mempool{}'s L1 memory is shared, a single \gls{dma} should access the entire L1 memory. Having multiple \glspl{dma} responsible for exclusive regions complicates programming. However, a single \gls{dma} connected to all \num{1024} memory banks implies excessive extra routing. A possible solution could be to reuse the L1 interconnect. However, this interconnect is not designed for wide data transfers, and a \gls{dma} could quickly congest the interconnect, slowing down the cores' transactions. In contrast, the \gls{axi} bus connecting all tiles is ideal for \gls{dma} type of transfers.

\begin{figure}[b]
  \centering
  \includegraphics[width=\linewidth]{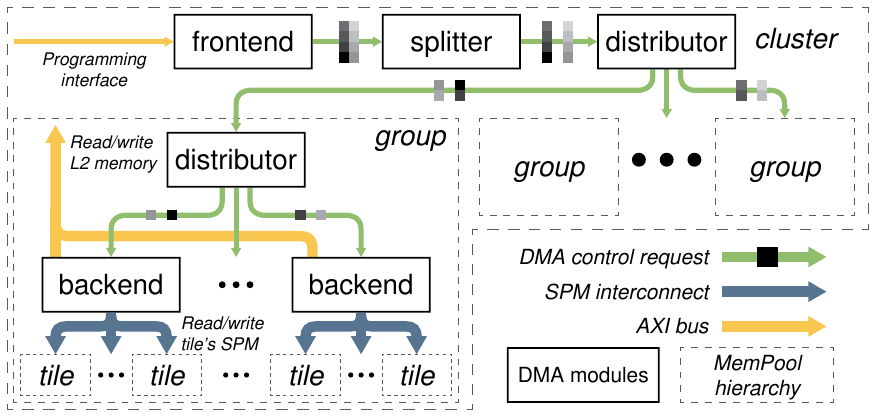}
  \caption{DMA architecture and its integration in \mempool{}. The \gls{dma} is programmed through a single \emph{frontend} which forwards a full \gls{dma} request to the \emph{splitter} module, where the request is split into multiple serial aligned requests. Through a tree of distributor modules, the requests are split into parallel, narrower requests, handled by distinct \emph{backends} that do the data movement. Each backend has access to L2 and to specific tiles.}
  \label{fig:dma_architecture}
\end{figure}

Using a modular \gls{dma} engine~\cite{Benz2023}, we design a distributed \gls{dma} for \mempool{} as shown in \cref{fig:dma_architecture}. It consists of a configurable number of data movers named \emph{backend}, e.g., one per four tiles, which are responsible for their tiles' memory region and connect to the tiles' \gls{axi} port on one side and their fully connected local crossbar on the other. A single configuration \emph{frontend} controls all backends. Specifically, the programmer can request \gls{dma} transfers involving the full \mempool{} cluster, and the \gls{dma} control will split the transfer into multiple smaller transfers and coordinate the data movers. To this end, we implement two key modules, a \emph{splitter} and a \emph{distributor}. Both take a \gls{dma} request as input and output one or multiple reshaped requests, which the data mover eventually executes. The former splits the transfer at the address boundary that spans one line of \mempool{}'s L1 memory into multiple serial requests, thereby taking the interleaved addressing scheme into account. The latter distributes those requests across multiple parallel requests requiring distinct L1 memory regions. This design has two main benefits. It fully reuses the hierarchical \gls{axi} interconnect to bring the data to the tiles with minimal extra routing. Also, the \glspl{dma} are connected to the tiles' internal, fully connected crossbars, as shown in \cref{fig:arch_overview}, where the bandwidth is high, and interference with the cores has the least impact.

%%%%%%%%%%%%%%%%%%%%%%%%%%%%%%%%%%%%%%%%%%%%%%%%%%%%%%%%%%%%%%%%%%%%%%%%%%%%%%%
%
% Full MemPool system
%
%%%%%%%%%%%%%%%%%%%%%%%%%%%%%%%%%%%%%%%%%%%%%%%%%%%%%%%%%%%%%%%%%%%%%%%%%%%%%%%

\subsection{System}\label{sec:system}

We connect the \gls{axi} ports to a \gls{soc} to get access to L2 (or main memory) as well as peripherals. \mempool{} also requires a few control registers for its runtime, which the \gls{soc} could integrate or be merged into the \mempool{} cluster itself. They allow waking up the cores, hold a few system parameters, such as the core count, and the \gls{ro} cache configuration.

We evaluate \mempool{} with a system containing a \SI{256}{\byte\per\cycle} L2 memory for instructions and data, the control registers to configure \mempool{} and send wake-up pulses, and a boot ROM. Each group has one \SI{512}{\bit} wide \gls{axi} bus connecting to the \gls{soc} with an access latency of 12 cycles.

\subsection{Evaluation}

We measure the execution time \changed{of the \matmul{} kernel described in \cref{subsec:microarch_benchmarking}} with a cold cache to evaluate various \gls{axi} interconnect architectures and \gls{ro} cache widths on the instruction path and compare it with a non-hierarchical, cacheless one. The \gls{ro} cache performs best if its line is larger or equal to the tiles' cache line width. A radix of eight gives the best performance with a speedup of 1.59$\times$. However, the associated hardware cost of three \gls{ro} caches is not justified compared to a radix-16 interconnect with only one \gls{ro} cache and a speedup of 1.54$\times$. Since the \gls{dma}'s performance is independent of the radix, we opt for the radix-16 solution.

\begin{figure}[b]
  \centering
  \includegraphics[width=\linewidth]{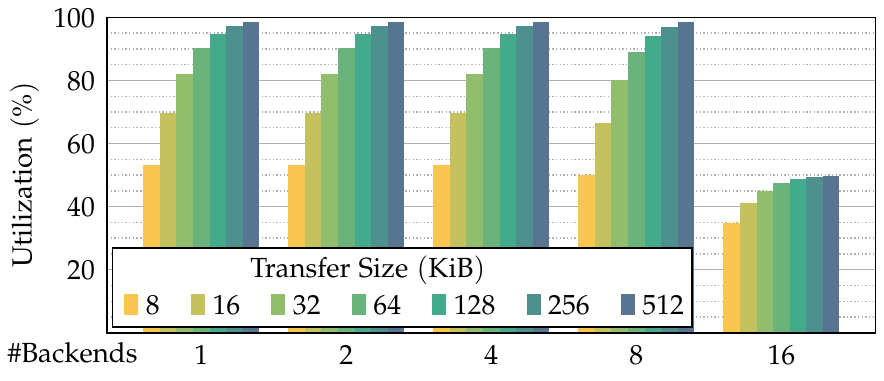}
  \caption{System bus utilization with different numbers of \gls{dma} \textit{Backends} per group for various transfer sizes.}
  \label{fig:dma_axi_results}
\end{figure}

To find the optimal configuration of the \gls{dma}, we measure the utilization of \mempool{}'s \gls{axi} master ports for various configurations and transfer sizes, as shown in \cref{fig:dma_axi_results}. Up to a specific size, the number of \gls{dma} backends makes little difference, and they can fully utilize the system bandwidth for large transfers. The utilization reaches roughly \SI{53}{\percent} even for very small transfers. Using 16 backends, one per tile, drastically reduces performance because each backend is only responsible for the \SI{512}{\bit} of continuous memory in a single tile preventing it from using \gls{axi} bursts. Our final implementation uses four \gls{dma} backends per group, as this configuration gives the best performance.

%%%%%%%%%%%%%%%%%%%%%%%%%%%%%%%%%%%%%%%%%%%%%%%%%%%%%%%%%%%%%%%%%%%%%%%%%%%%%%%
%
% Physical implementation
%
%%%%%%%%%%%%%%%%%%%%%%%%%%%%%%%%%%%%%%%%%%%%%%%%%%%%%%%%%%%%%%%%%%%%%%%%%%%%%%%

\section{Implementation}\label{sec:implementation}

We have individually explored and evaluated all the critical components of \mempool{}. In this section, we analyze the full \mempool{} design, describing the implementation methodology, \mempool{}'s final parameterization, and its physical implementation used for the final evaluation.

\subsection{Methodology}\label{sec:methodology}

\mempool{} targets running at \SI{500}{\mega\hertz} in worst-case conditions (SS/\SI{0.72}{\volt}\kern-.15em/\SI{125}{\celsius}) when implemented in GlobalFoundries' 22FDX \gls{fdsoi} technology. We implement \mempool{} using Synopsys' Fusion Compiler 2022.03. Due to the size of the design, we adopt a bottom-up implementation flow, first implementing the groups before integrating them as macros into the cluster design. We use Synopsys' PrimeTime 2022.03 to estimate \mempool{}'s power consumption in typical conditions (TT/\SI{0.80}{\volt}\kern-.16em/\SI{25}{\celsius}), with switching activities extracted from a post-layout gate-level simulation running at \SI{600}{\mega\hertz}. All performance measurements come from cycle-accurate \gls{rtl} simulation.

\subsection{Implementation flow}

Routing \mempool{} is a fundamental challenge of its physical implementation due to its large size and the need for a hierarchical approach. The chosen \topology{H} interconnect introduces a group hierarchy, allowing us to bypass the tile hierarchy and implement a flat group as the first level. This allows the tools to implement all interconnects flat and route through the tiles, resulting in improved resource utilization and a smaller footprint compared to previous work~\cite{Cavalcante2021}. Despite the addition of custom \gls{dsp} units, a full \gls{axi} interconnect, cache hierarchy, and \gls{dma}, we achieve the same worst-case frequency of \SI{482}{\mega\hertz} while reducing the area by 1.6$\times$ to \SI{12.8}{\milli\meter\squared}, thanks to the optimized cache and improved implementation flow. The deepest path traverses 35 gates in the \gls{ipu}, but the interconnect connecting two groups actually limits \mempool{}'s frequency. This path is \SI{40}{\percent} wire delay and \SI{60}{\percent} gate delay (28 gates), half of which are buffers.

\subsection{Floorplan}

\begin{figure}[bt]
  \centering
  \includegraphics[width=\linewidth]{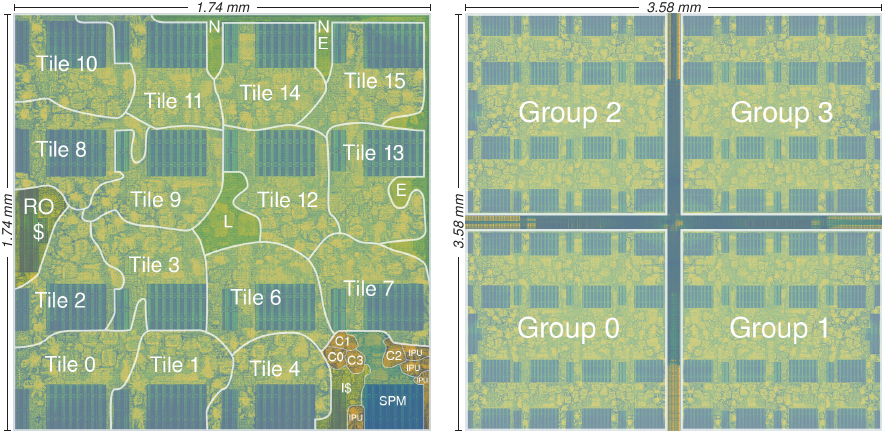}
  \caption{Annotated dieshot of placed and routed \mempool{} group (left) and cluster (right). We highlight the tiles, the \gls{ro} cache, and the interconnects between tiles of the same group (L) and to other groups (N, NE, E) as well as Tile 5's cores, \glspl{ipu}, instruction cache, and \gls{spm} with its interconnect. }
  \label{fig:dieshot_group}
\end{figure}

The annotated die shot of the fully placed and routed \mempool{} group can be seen in \cref{fig:dieshot_group}. We place the macros of the tile memories in a \by{4}{4} grid, flipping the upper half of the grid to create an open area in the middle of the group, where the tool has space to place the interconnects. The annotations show how all the tiles are clustered around their memory macros but drawn to the center to minimize wiring delay towards the interconnect. The remote \gls{spm} interconnects are drawn to their pins and fit between the tiles. The \gls{axi} interconnect and \gls{ro} cache are placed on the left boundary.

\begin{figure}[bt]
  \centering
  \includegraphics[width=\linewidth]{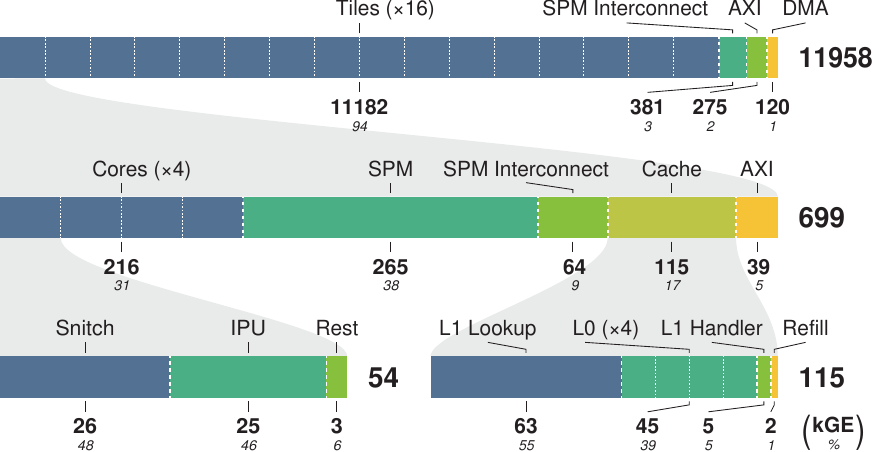}
  \caption{Hierarchical area breakdown of a \mempool{} group in \textbf{\si{\kilo\ge}} with annotations showing the \textit{percentage} of the immediate parent component. The whole group occupies \SI{12}{\mega\ge}, most of which is occupied by tiles, while the interconnects and \glspl{dma} only make up a small percentage. The cores and \gls{spm} banks comprise most of the area within a tile.}
  \label{fig:area_breakdown}
\end{figure}

\cref{fig:area_breakdown} shows the hierarchical area distribution of \mempool{}'s group. It shows that we succeeded in adding an \gls{axi} interconnect with minimal area overhead, despite including a full level of cache hierarchy and a specialized \gls{dma}. Within a tile, the \gls{spm} banks are the biggest components, followed by the four cores and their instruction cache. The core's area is split between the Snitch core itself and its \gls{ipu}.

%%%%%%%%%%%%%%%%%%%%%%%%%%%%%%%%%%%%%%%%%%%%%%%%%%%%%%%%%%%%%%%%%%%%%%%%%%%%%%%
%
% Programming Model
%
%%%%%%%%%%%%%%%%%%%%%%%%%%%%%%%%%%%%%%%%%%%%%%%%%%%%%%%%%%%%%%%%%%%%%%%%%%%%%%%

\section{Programming Model}\label{sec:programming_model}

A streamlined and efficient programming model is vital for adopting a new computing platform. We have designed \mempool{} with programmability as one of its primary goals.\footnote{To facilitate developing applications for \mempool{}, we support emulating it on Banshee, a binary-translation-based emulation platform~\cite{Riedel2021}.}

\mempool{}'s shared memory already simplifies programming compared to designs with isolated memory regions, such as multi-cluster or systolic architectures. It enables all-to-all communication and synchronization without considering the memory map. Furthermore, the independent and individually programmable cores eliminate problems such as lock-step execution, branch divergence, or thread warping.

\subsection{Compiler support}
We have extended the GNU GCC and the LLVM toolchains to support \mempool{}'s \gls{isa} extensions and instruction scheduling. Both toolchains are aware of \mempool{}'s architectural latencies and will schedule instructions to prevent \gls{raw} hazards. This entails scheduling load operations as far as possible from their usage to hide the L1 latency. Similarly, pipelined instructions offloaded to Snitch's accelerator are scheduled. Instruction scheduling in the compiler allows hiding the non-idealities of the L1 interconnect and Snitch's accelerator from the programmer. For them, \mempool{} becomes an idealized single-cycle latency cluster.

\subsection{Synchronization}
\changed{A key aspect in programming a manycore system is synchronization. Consistent with its \riscv{} cores, \mempool{} follows the \gls{rvwmo} memory model and relies on the \riscv{} standard atomic extension for synchronization. Specifically, each L1 \gls{spm} bank's controller is extended with a small \gls{alu} to support \riscv{}'s \glspl{amo} and \gls{lrsc} instructions. For the latter, the memory controller contains a reservation register where a \gls{lr} can place a reservation for an address. This reservation is valid until the memory location changes and determines the outcome of the \gls{sc}. They allow the implementation of mutexes or concurrent algorithms in a standardized fashion. To support the \gls{rvwmo} model, Snitch implements \riscv{}'s fence instructions that can be used to order memory transactions between cores. To save energy during synchronization, \mempool{} also provides sleep and wake-up behavior, where each core can send a wake-up pulse to arbitrary cores or wake up the complete clusters in a single store, enabling very efficient synchronization barriers.}

\subsection{Runtimes}

\subsubsection{Bare-metal runtime}
We provide a bare-metal C runtime for \mempool{} where cores are programmed with the same code, and the programmer can create branches for each core or let them work on specific data using their unique ID~\cite{Tagliavini2018}. We hand control over to the programmer once the stack is allocated in the sequential region and the runtime is initialized. The programmer can then use various runtime functions, such as allocators and barriers, and has complete control over every core to implement highly optimized applications. However, implementing large applications becomes time-consuming.

\subsubsection{OpenMP}
We implement the standard OpenMP runtime for \mempool{} in C, providing a shared-memory fork-join programming model. \changed{The program is primarily executed by a single \emph{master} core, which can then fork off to employ the cores waiting for work through wake-up triggers and synchronization with \riscv{} atomics.} We support static and dynamic loop scheduling; parallel sections; synchronization directives such as master, critical, atomic, and barrier; and reductions. OpenMP simplifies parallelizing a workload at the cost of runtime overhead, as evaluated in \cref{subsec:system_benchmarking_fullapps}.

\subsubsection{Halide}
Halide is a \gls{dsl} aimed at image processing and machine learning~\cite{Ragan-Kelley2012}. It decouples the functional description of an application from its execution, such as tiling, unrolling, or parallelizing, allowing sharing of functional code across different platforms while tuning the implementation for each architecture. Its functional programming nature allows Halide to optimize across kernel boundaries. \changed{We extend Halide to support \mempool{} by enabling \riscv{} support in its LLVM-based backend. We implement Halide's runtime in C, most importantly, fork/join functions to support the \emph{parallel} schedule and dynamic memory management to create temporary buffers. The remaining schedules (unrolling, tiling, etc.) are natively supported through Halide's LLVM backend.}

%%%%%%%%%%%%%%%%%%%%%%%%%%%%%%%%%%%%%%%%%%%%%%%%%%%%%%%%%%%%%%%%%%%%%%%%%%%%%%%
%
% Results
%
%%%%%%%%%%%%%%%%%%%%%%%%%%%%%%%%%%%%%%%%%%%%%%%%%%%%%%%%%%%%%%%%%%%%%%%%%%%%%%%

\section{Performance Evaluation}\label{sec:evaluation}

\subsection{Microarchitecture Benchmarking}\label{subsec:microarch_benchmarking}

We evaluate \mempool{}'s performance, power, and energy efficiency for \gls{dsp} kernels from a wide range of domains. All kernels operate on 32-bit integers, are parallelized across all of \mempool{}'s cores, and include a final synchronization barrier. Specifically, we consider the following kernels:

\setlist[description]{font=\normalfont}
\begin{description}

\item[\matmul{}:]
A matrix-matrix multiplication where each core operates on a $4 \times 4$ output tile to maximize computational intensity, resulting in eight loads per 16 \gls{mac} operations.

\item[\conv{}:]
A 2D convolution with a $3 \times 3$ kernel. Cores operate on pixels mapped to their tile, leading to local accesses except for pixels at the edges of a tile. The kernel maximizes data reuse by working on $4 \times 3$ tiles.

\item[\dct{}:]
Computes the 2D \acrfull{dct} on $8 \times 8$ blocks, as used for JPEG compression. Cores work on local blocks and use the stack for intermediate results.

\item[\axpy{}:]
A key BLAS routine computing $\alpha \cdot \vec{x} + \vec{y}$ with a low computational intensity with two loads and one store for every \gls{mac}. It is optimized only to have local accesses.

\item[\dotp{}:]
The dot product computes the scalar product of two vectors. Like \axpy{}, it has a low computational intensity and is parallelized only to have local accesses.

\end{description}

\subsubsection{Performance Results}\label{subsec:microarch_benchmarking_results}

\Cref{tab:perfcomp} shows the performance metrics of the selected kernels. All results were extracted with the methodology described in \cref{sec:methodology}. The full \mempool{} cluster consumes roughly \SI{1.5}{\watt} and consistently achieves a high \gls{ipc}, especially for compute-intensive kernels, where we can reach up to \SI{336}{\op\per\cycle} thanks to the \gls{mac} extension. In terms of energy-efficiency, \mempool{} reaches \SI{159}{\giga\ops\per\watt}.

\begin{table}[tb]
  \centering
  \caption{Benchmark and power results obtained from post-layout simulations. An operation corresponds to a 32-bit addition or multiplication.}
  \label{tab:perfcomp}
  \vspace{-0.2cm}
  \begin{tabular}{@{}lcrrrr@{}}
    \toprule
    \multirow{2}{*}{Kernel} & \multirow{2}{*}{Size} & Util & Power &  \multicolumn{2}{c}{Performance} \\
    &  & $\left(\text{\si{\ipc}}\right)$ & $\left(\text{\si{\watt}}\right)$ & $\left(\text{\si{\op\per\cycle}}\right)$ & $\left(\text{\si{\giga\ops\per\watt}}\right)$ \\
    \midrule
    \matmul{} & $256\times256$  & 0.88 & \num{1.67} & 285 &  103 \\
    \conv{}   & $96\times1024$  & 0.87 & \num{1.27} & 336 &  159 \\
    \dct{}    & $192\times1024$ & 0.93 & \num{1.09} & 168 &   92 \\
    \axpy{}   & $98'304$        & 0.76 & \num{1.51} &  90 &   36 \\
    \dotp{}   & $98'304$        & 0.74 & \num{1.50} &  92 &   37 \\
    \bottomrule
  \end{tabular}
\end{table}

\subsubsection{Scaling behavior}

\begin{figure}[bt]
  \centering
  \includegraphics[width=\linewidth]{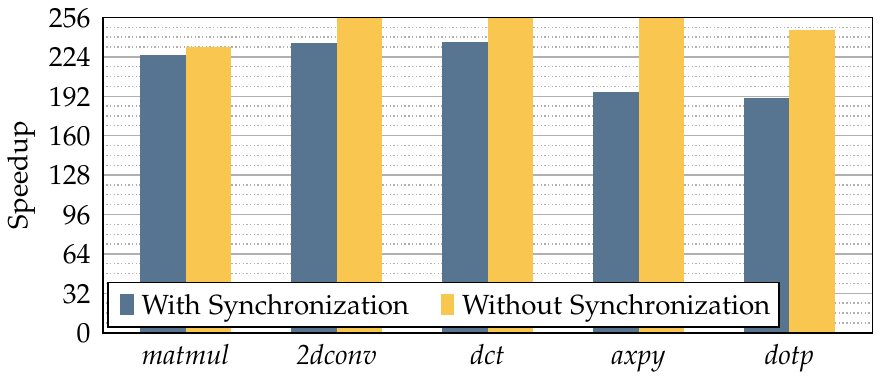}
  \caption{Speedup of \mempool{} over single-core execution using weak scaling, i.e., the problem scales with the core count. We show the speedup with and without a final synchronization barrier to analyze the main cause of performance losses.}
  \label{fig:eval_scaling}
\end{figure}

A comparison with an idealized, conflict-free, single-core system illustrates how well \mempool{} scales. We use weak scaling, which means the problem size scales with the system size. \Cref{fig:eval_scaling} shows \mempool{}'s speedup with and without a full final synchronization barrier. It allows analyzing whether the inherent final synchronization step or conflicts between cores limit scaling. For compute-intensive kernels such as \matmul{}, \conv{}, and \dct{}, the speedup is very close to ideal, even with the final barrier. While \conv{} and \dct{}'s performance, excluding the synchronization, is on par with the idealized system, \matmul{}'s speedup indicates that conflicts lead to performance losses. For kernels with low compute intensity, the final synchronization step becomes noticeable. However, they still achieve \SI{75}{\percent} of the ideal speedup. Overall, \mempool{} achieves very close to ideal speedups, only losing \SI{10}{\percent} due to synchronization in compute-intensive workloads.

We further investigate \mempool{}'s non-idealities by analyzing where the cores spend their time. Some cycles need to be spent on control and memory instructions due to Snitch's simple design and \riscv{}'s load-store architecture, inhibiting a \SI{100}{\percent} compute unit utilization. \Cref{fig:eval_kernel_stalls} shows the breakdown of cycles spent. \changed{It stacks the percentage of compute instructions (operations counted in the kernel's arithmetic intensity), representing compute unit utilization, with the percentage of control instructions (memory operations, address increments, control flow instructions), totaling in the \gls{ipc}}. Finally, it shows the breakdown of idle cycles due to synchronization or architectural stalls. A clear distinction between compute and memory-intensive kernels can be made out. The former achieve a high compute utilization of up to \SI{66}{\percent}, while the latter spend more time on load and store instructions inherent to load-store architectures.

\begin{figure}[tb]
  \centering
  \includegraphics[width=\linewidth]{fig/kernel_stalls}
  \caption{Breakdown of cores' activity during kernel execution. The first two bars show the time spent executing instructions, \changed{differentiating between \emph{compute} (additions, multiplication, etc.) and \emph{control} (loads, address increments, branches, etc.) instructions.} All other bars indicate idle phases. \emph{Synchronization} represents the time spent sleeping at barriers, \emph{I\$} denotes instruction path stalls, \emph{LSU} stalls come from congestion in the interconnect preventing the core from issuing a load or store, and \emph{\gls{raw}} stalls denote \acrlong{raw} hazards. It contains metrics such as the compute unit utilization, the \gls{ipc}, and the synchronization costs.}
  \label{fig:eval_kernel_stalls}
\end{figure}

Another difference between compute and memory-bound kernels is the synchronization overhead. While the absolute overhead is similar, compute-intensive kernels amortize it by running longer on the same amount of data. We observe very few architectural stalls. The only kernel with \emph{LSU} stalls, which come from conflicts in the interconnect, is \matmul{}. The hybrid addressing scheme keeps most of the other kernels' data local to minimize memory conflicts. Only the \dotp{}'s reduction step exhibits some conflicts. \mempool{}'s low-latency interconnect and Snitch's ability to manage multiple outstanding operations minimizes the \gls{raw} stalls to a negligible amount across all kernels. Finally, the well-optimized cache and its prefetching lead to very few instruction stalls.

\subsection{System Benchmarking}\label{subsec:system_benchmarking}

\subsubsection{Double-buffered implementation}\label{subsec:system_benchmarking_doublebuffered}

We evaluate the performance of the full \mempool{} with double-buffered kernels operating on data from system memory. Because the L1 has to hold two problems simultaneously, the kernels work on half the problem sizes here. The \glspl{pe} synchronize between rounds, where the first \gls{pe} entering a new round checks whether the \gls{dma} transfer has finished. The last \gls{pe} programs the \gls{dma} transfer for the next round.

\begin{figure}[b]
  \centering
  \includegraphics[]{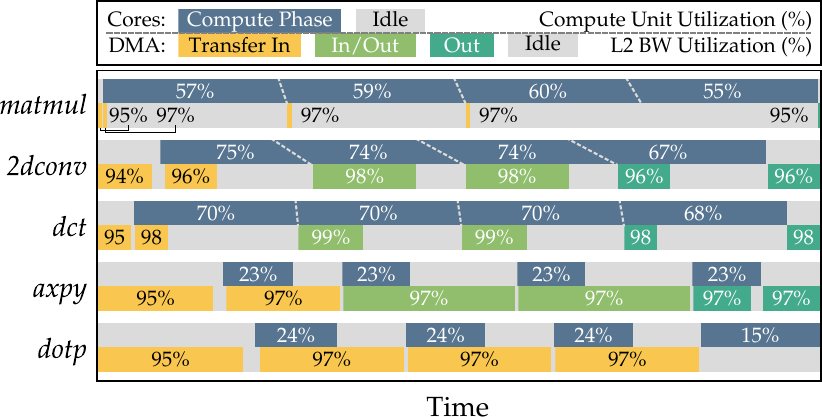}
  \caption{Timing diagram of double-buffered kernels. The upper bar shows the active computing phases and compute unit utilization. The lower bar represents the \gls{dma} phases and utilization. Diagonal lines mark fused phases, where some \glspl{pe} already move on to the next phase.}
  \label{fig:eval_double_buffer}
\end{figure}

\cref{fig:eval_double_buffer} shows the compute and transfer phases, starting with an initial \gls{dma}-only phase loading in the first chunk, followed by the first compute phase, which is accompanied by a transfer loading the second chunk. The next two rounds are full compute and transfer rounds, where previous results are stored back and upcoming inputs transferred in. These stages are the most important as they are replicated for large problem sizes while the ramp-up and down rounds stay constant. A transfer with only outputs accompanies the last compute round before a final \gls{dma}-only phase writes back the last results. It takes roughly 30 cycles to set up a new \gls{dma} transfer, and each L2 access has a latency of 12 cycles.

Compute-bound kernels, like \matmul{}, \conv{}, and \dct{}, achieve even higher performance during steady rounds than with a single compute round. We achieve an \gls{ipc} of \SI{94}{\percent} or \SI{306}{\op\per\cycle} for \matmul{}, \SI{0.99}{\ipc} and \SI{178}{\op\per\cycle} for \dct{}, or even \SI{0.98}{\ipc} and \SI{381}{\op\per\cycle} for \conv{}, which corresponds to \SI{229}{\giga\ops} or \SI{180}{\giga\ops\per\watt}. This speedup is mainly due to fusing compute rounds and smaller synchronization overheads. The fused rounds also eliminate most of the LSU stalls, which mainly originated from cores initially accessing the same banks. The slight drifting of cores and their access patterns leads to more uniformly distributed accesses and fewer conflicts. Here, the cores keep this drift across compute phases and suffer fewer LSU stalls. Finally, for heavily memory-bound kernels like \axpy{} and \dotp{}, the L2 bandwidth cannot keep all \glspl{pe} occupied. While they achieve \glspl{ipc} of \SI{97}{\percent} and \SI{99}{\percent}, respectively, the compute phases only last for \SI{35}{\percent} and \SI{51}{\percent} of the steady state rounds. Due to the reduction and writeback, \dotp{}'s last compute phase is longer.

\subsubsection{OpenMP and Halide applications}\label{subsec:system_benchmarking_fullapps}

We have shown that \mempool{} can achieve very high performance on well-optimized \gls{dsp} kernels. While hand-optimized kernels are often used in applications through libraries, we also show that \mempool{} is conveniently programmable with high-level abstractions such as OpenMP and the \gls{dsl} Halide with the following applications:

\setlist[description]{font=\normalfont}
\begin{description}

\item[\histequal{}:]
To enhance an image's contrast, histogram equalization transforms an image based on its histogram such that the output image uses the full spectrum of intensities. We implement it in Halide, where the main challenge lies in synchronization since this algorithm has multiple reduction and serial steps.

\item[\raytracing{}:]
A standard method to render photo-realistic 3D scenes is ray tracing. We implement a basic integer-based version in C and parallelize it with OpenMP. While each pixel can be rendered independently, the workload balancing is challenging because ray tracing is non-data-oblivious, meaning its execution depends on the input data. Specifically, the time to render each ray depends on the objects in its path. It shows the benefit of individually programmable cores and the flexibility of \mempool{}.

\changed{
\item[\textit{\gls{bfs}}:]
A graph traversal algorithm that visits all vertices level by level. It uses a queue to store the current level's vertex and keeps track of visited vertices to avoid revisiting them. The algorithm is parallelized with OpenMP. All \glspl{pe} search for new vertices in parallel and atomically update the list of visited vertices and the queue of vertices for the next level. A barrier synchronizes all cores between levels.
}

\end{description}

All algorithms run on a single-core and the full \mempool{} to demonstrate that \mempool{} can efficiently run complete applications with minimal parallelization overhead.

The \histequal{} application achieves \SI{40}{\percent} of the linear speedup and is mainly limited by its sequential parts and reductions, such as the computation of the transformation function. This speedup is very close to the ideal speedup considering Amdahl's law, illustrating how \mempool{} can parallelize complex algorithms with little overhead.

Ray tracing poses different challenges. It is fully parallelizable, but balancing the workload is challenging. Nevertheless, \mempool{} achieves \SI{91}{\percent} of the ideal speedup. \SI{3}{\percent} of the loss in performance comes from the imbalance in the workload distribution, while the remaining \SI{6}{\percent} can be found in the runtime overhead of OpenMP's dynamic scheduling. 

\changed{
The \bfs{} algorithm is inherently hard to parallelize and requires many shared data structures to be accessed atomically. Furthermore, the vertices-per-level vary heavily, making workload balancing difficult. Nevertheless, \mempool{} achieves \SI{51}{\percent} of the ideal speedup. The additional instructions required to access the shared data structures cost \SI{32}{\percent}, while \SI{17}{\percent} is lost due to unbalanced workloads.
}

These applications show how \mempool{} enables implementing and running complex and data-dependent kernels with high-level abstractions. Its shared memory and independent cores simplify parallelizing applications efficiently.

\subsubsection{Energy of individual instructions}\label{subsec:power_system_benchmarking_fullapps}

\begin{figure}[bt]
  \centering
  \includegraphics[width=\linewidth]{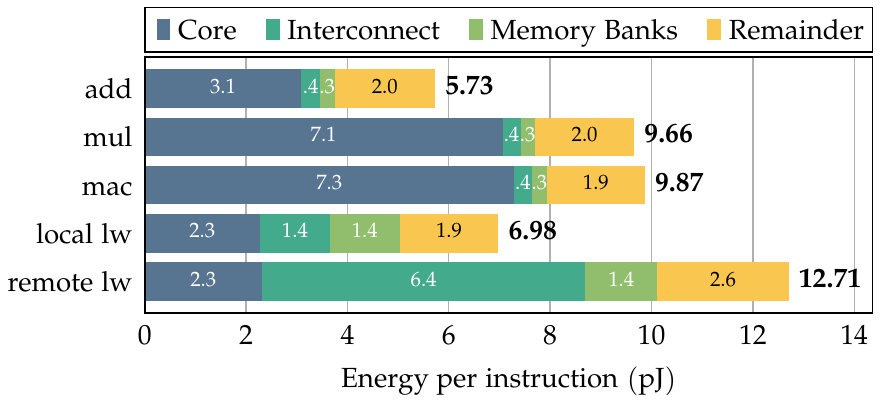}
  \caption{Energy breakdown of individual instructions extracted from post-layout simulations. We measure the energy of three different arithmetic instructions and the load word instruction when loading from banks in the same (\emph{local}) tile or \emph{remote} ones, thereby passing through the groups. The \emph{Remainder} category includes all components with minimal contributions, mainly leakage, like the caches, \gls{dma}, \gls{axi}, etc.}
  \label{fig:eval_instr_energy}
\end{figure}

The energy consumed by different instructions quantifies the benefits of our \gls{isa} extensions and hybrid addressing scheme. \Cref{fig:eval_instr_energy} shows the energy breakdown, including dynamic and static power, while all cores execute the instruction with randomized data. We report the energy per core per cycle since all instructions, including multi-cycle ones, are fully pipelined, and we execute one instruction per cycle per core.

The arithmetic instructions' energy consumption motivates our \gls{isa} supporting instructions like the \gls{mac}. Fusing the addition and multiplication increases a multiplications energy consumption by only \SI{0.2}{\pico\joule}, but eliminates the need for the add instruction, thereby saving \SI{36}{\percent}.

Similarly, the energy of local and remote memory transactions emphasizes the benefit of our hybrid addressing scheme. A remote transaction consumes 1.8$\times$ the energy of a local one. Therefore, the hybrid addressing scheme improves energy efficiency by reducing the number of remote accesses. Comparing remote loads to arithmetic instructions shows how efficient \mempool{}'s interconnect design is. Crossing the whole \mempool{} macro takes only \SI{29}{\percent} more energy as a \gls{mac} operation, thanks to the pipelined interconnect. Our interconnect remains energy efficient, despite being scaled to 256 initiators and 1024 targets.

\subsubsection{Power breakdown}\label{subsec:breakdown_system_benchmarking_fullapps}

\begin{figure}[bth]
  \centering
  \includegraphics[width=\linewidth]{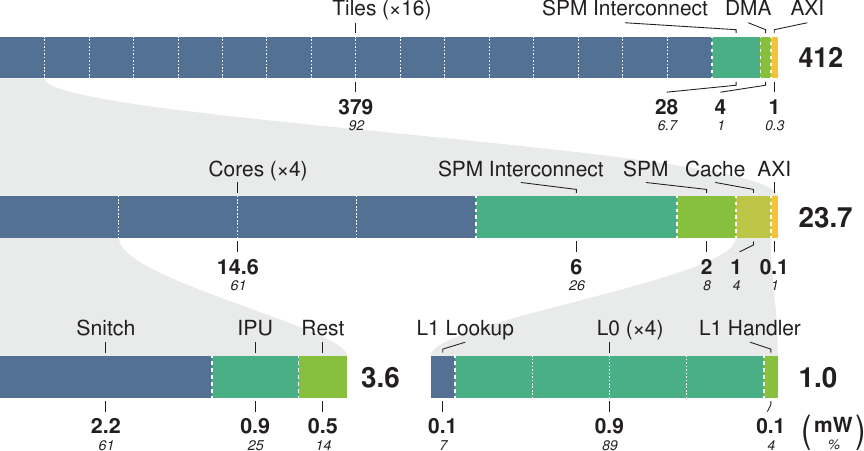}
  \caption{Hierarchical power breakdown of \mempool{} executing a \matmul{} kernel. The results are shown in \textbf{\si{\milli\watt}} and \textit{percentage} and are extracted using PrimeTime with annotations from a post-layout simulation.}
  \label{fig:eval_powerbreakdown}
\end{figure}

To gain deeper insights into \mempool{}'s power consumption, \cref{fig:eval_powerbreakdown} breaks the power consumption down into individual components. Despite scaling the cluster to 256 cores with low-latency access to L1, most of the power is still spent on computation. Specifically, \SI{56}{\percent} of the power is consumed by the cores, while the contribution of the \gls{spm} interconnect is kept at \SI{30}{\percent} despite its size. The memory banks themselves account for \SI{7}{\percent}.

%%%%%%%%%%%%%%%%%%%%%%%%%%%%%%%%%%%%%%%%%%%%%%%%%%%%%%%%%%%%%%%%%%%%%%%%%%%%%%%
%
% Related Work
%
%%%%%%%%%%%%%%%%%%%%%%%%%%%%%%%%%%%%%%%%%%%%%%%%%%%%%%%%%%%%%%%%%%%%%%%%%%%%%%%

\glsreset{sm}

\section{Related Work}\label{sec:related_work}

Energy-efficient manycore systems have been studied extensively, and various architectural approaches exist to tackle scaling parallel systems \cite{Muralidhar2022}. In the following, we focus on manycore systems based on clusters of \glspl{pe}. \changed{Because fair quantitative comparisons between different designs are challenging due to numerous design variables and differences in technology, core counts, and benchmarks, we provide a qualitative comparison of existing systems in \cref{tab:related_work}.}

%Single-cluster systems not scaled
A cluster of \glspl{pe} sharing a low-latency L1 memory is a typical architecture for efficiently tackling today's parallel workloads. For example, Greenwaves' GAP9 processor's~\cite{GreenWavesTechnologies2021} compute power comes from nine independent \riscv{} cores coupled to a shared L1 \gls{spm}. Its primary architectural idea is very similar to \mempool{}. However, \mempool{} presents a way to scale this idea to 256 cores.
%Single-cluster systems scaled? None, except MemPool?
Ramon Chip's RC64~\cite{Ginosar2016} targets task parallelism and implements a similar architecture to \mempool{}, coupling 64 \gls{dsp} cores to a shared \gls{spm} accessible in a small number of cycles. Each core is equipped with a private instruction and data cache and a private \gls{spm} to reduce bandwidth requirements to the shared \gls{spm}. \mempool{} scales the single-cluster design even further without relying on costly data caches to reduce bandwidth requirements.

%Multicluster systems build from many single-cluster systems
Other architectures rely on replicating the small compute cluster to scale to hundreds of cores. Manticore~\cite{Zaruba2021} is built out of a similar compute cluster as GAP9, coupling eight small \SIadj{32}{\bit} \riscv{} cores to a low-latency shared L1 memory. This cluster is replicated and connected to a shared L2 memory via an \gls{axi} interconnect to build a system of 1024 \glspl{pe} per chiplet. Kalray's MPPS3-80~\cite{DupontdeDinechin2021} follows a similar approach with larger cores. 16 \SIadj{64}{\bit} processors with private L1 data caches form a cluster with 4MB of L2 \gls{spm} memory. The chip features five clusters, resulting in 80 cores. Esperanto's ET-SoC-1~\cite{Ditzel2021} has a similar hierarchy and has \num{1088} \SIadj{64}{\bit} \riscv{} cores organized in \num{34} \emph{shires} of 32 cores and \SI{4}{\mebi\byte} of \gls{spm} each. While this approach allows scaling into the manycore regime, it creates many small clusters with private memories, complicating the programming model and incurring communication overhead between clusters.

\colorlet{coloryes}{OliveGreen}
\colorlet{colorjein}{YellowOrange}
\colorlet{colorno}{OrangeRed}

\colorlet{colorverygood}{OliveGreen}
\colorlet{colorgood}{OliveGreen}
\colorlet{colorneutral}{Gray}
\colorlet{colorbad}{OrangeRed}
\colorlet{colorverybad}{OrangeRed}

\newcommand{\verygood}[1]{\textcolor{colorverygood}{#1}}
\newcommand{\good}[1]{\textcolor{colorgood}{#1}}
\newcommand{\neutral}[1]{\textcolor{colorneutral}{#1}}
\newcommand{\bad}[1]{\textcolor{colorbad}{#1}}
\newcommand{\verybad}[1]{\textcolor{colorverybad}{#1}}

\newcommand\circledsym[2]{%
  \adjustbox{height=1.15em,margin*=0 -1.25 -2.5 0}{%
    \tikz\node[circle,color=white,fill=#1,inner sep=.2pt,font=\bfseries]{#2};%
  }
}

\newcommand{\yes}{\circledsym{coloryes}{$\pmb\checkmark$}}
\newcommand{\jein}{\circledsym{colorjein}{$\pmb\approx$}}
\newcommand{\no}{\circledsym{colorno}{$\pmb\times$}}
\newcommand{\unknown}{\circledsym{colorneutral}{?}}

\newcolumntype{Y}{>{\centering\arraybackslash}X}
\newcolumntype{Z}{>{\raggedleft\arraybackslash}X}
\newcolumntype{R}{%
  >{\adjustbox{right=6em,angle=310,lap=-\width+1.5em}\bgroup}%
  c%
  <{\egroup}%
}
\newcolumntype{M}{%
  >{\adjustbox{right=6em,angle=310,lap=-\width+2.5em}\bgroup}%
  c%
  <{\egroup}%
}
\newcommand*\rot{\multicolumn{1}{R}}%
\newcommand*\rotm{\multicolumn{1}{M}}%
\newcommand{\tblrottitle}[1]{\rot{\textbf{#1}}}
\newcommand{\tblrottitlem}[1]{\rotm{\textbf{\makecell[cr]{#1}}}}
\newcommand{\productentry}[2]{#1~\cite{#2}}

\newcommand{\underlinecenter}[2]{%
\setul{3pt}{.4pt}% 5pt below contents
\ul{\mbox{\hspace{#1}}#2\mbox{\hspace{#1}}}}

\begin{table}[tb]
  \caption{Comparison of \mempool{} with related, cluster-based architectures.}
  \label{tab:related_work}
  \vspace{-0.4cm} % reduce distance above table
  % \scriptsize%
  \setlength{\tabcolsep}{1pt}%
  \sisetup{range-phrase=--}%
  \center%
  \begin{tabularx}{\columnwidth}{@{}lllcZYYY@{}}
    \toprule
     &
    \parbox[t][5em][b]{1em}{\textbf{Architecture}}
      & \parbox[t][5em][b]{1em}{\textbf{ISA}}
      & \tblrottitle{Cluster \glspl{pe}}
      & \tblrottitle{Total \glspl{pe}}
      & \tblrottitlem{Independent \\ \glspl{pe}}
      & \tblrottitlem{Low latency \\ interconnect}
      & \tblrottitle{Open source}
      \\
    \midrule
    \parbox[t]{4mm}{\multirow{7}{*}{\rotatebox[origin=c]{90}{\underlinecenter{2.15em}{\textit{mesh-based}}}}} &
    \productentry{RAW}{Taylor2002}
      & \neutral{32-bit MIPS-style}
      & \bad{-}
      & \bad{16}
      & \yes{}
      & \no{} % Hops but max 6 cycles from corner to corner. However, also 6 cycles back...
      & \no{}
      \\
      &
    \productentry{Celerity}{Davidson2018}
      & \neutral{32-bit RISC-V}
      & \bad{-}
      & \good{496}\textsuperscript{*}
      & \yes{}
      & \no{}
      & \yes{}
      \\
      &
    \productentry{KiloCore}{Bohnenstiehl2017}
      & \neutral{40-bit RISC}
      & \bad{-}
      & \good{1000}
      & \yes{}
      & \no{}
      & \no{}
      \\
      &
    \productentry{Piton}{McKeown2017}
      & \neutral{64-bit SPARC V9}
      & \bad{-}
      & \bad{25}
      & \yes{}
      & \no{}
      & \yes{}
      \\
      &
    \productentry{TILE64}{Bell2008}
      & \neutral{64-bit VLIW}
      & \bad{-}
      & \neutral{64}
      & \yes{}
      & \no{}
      & \no{}
      \\
      &
    \productentry{Epiphany-V}{Olofsson2015}
      & \neutral{64-bit RISC}
      & \bad{-}
      & \good{1024}
      & \yes{}
      & \no{}
      & \no{}
      \\
      &
    \productentry{Pixel Visual Core}{Redgrave2018}
      & \neutral{16-bit VLIW}
      & \good{256}
      & \good{2048}
      & \no{}
      & \no{}
      & \no{}
      \\
    \parbox[t]{4mm}{\multirow{7}{*}{\rotatebox[origin=c]{90}{\underlinecenter{1.55em}{\textit{crossbar-based}}}}} &
    \productentry{GAP9}{GreenWavesTechnologies2021}
      & \neutral{32-bit \riscv{}}
      & \verybad{9}
      & \verybad{9}
      & \yes{}
      & \yes{}
      & \jein{}
      \\
      &
    \productentry{RC64}{Ginosar2016}
      & \neutral{32-bit VLIW}
      & \good{64}
      & \neutral{64}
      & \yes{}
      & \yes{}
      & \no{}
      \\
      &
    \productentry{Manticore}{Zaruba2021}
      & \neutral{32-bit \riscv{}}
      & \bad{8}
      & \good{4096}
      & \yes{}
      & \no{}
      & \yes{}
      \\
      &
    \productentry{MPPA3}{DupontdeDinechin2021}
      & \neutral{64-bit VLIW}
      & \neutral{16}
      & \neutral{80}
      & \yes{}
      & \no{}
      & \no{}
      \\
      &
    \productentry{ET-SoC-1}{Ditzel2021}
      & \neutral{64-bit \riscv{}}
      & \good{32}
      & \good{1088}
      & \yes{}
      & \no{}
      & \no{}
      \\
      &
    \productentry{H1000}{NVIDIACorporation2022}
      & \neutral{32/64-bit PTX}
      & \good{128}
      & \good{18432}
      & \no{}
      & \no{}
      & \no{}
      \\
      &
    \textbf{This Work}
      & \neutral{32-bit \riscv{}}
      & \good{256}
      & \good{256}
      & \yes{}
      & \yes{}
      & \yes{}
      \\
    \bottomrule
  \end{tabularx}
  \scriptsize
  \raggedright
  \raisebox{-.2ex}{\jein{}}~=~Closed source based on open source.
  *Contains additional five Linux-capable 64-bit cores and ten ultra-low-power cores.
\end{table}

%GPU
Modern \glspl{gpu} resemble multi-cluster designs, with \glspl{sm} corresponding to shared-memory clusters that act as basic building blocks. NVIDIA's leading-edge H100~\cite{NVIDIACorporation2022} features 144 \glspl{sm}, each featuring 128 FP32 units. However, \glspl{gpu} operate in the \gls{simt} regime, meaning that a single instruction stream controls multiple compute units. Each \gls{sm} has four scheduler and dispatch units controlling 32 FP32 units each, resulting in four independent instruction streams per \gls{sm} or 576 per \gls{gpu}. \Glspl{gpu} feature thousands of compute units, but their utilization is limited by the \gls{simt} regime. Especially for irregular and non-data-oblivious algorithms, thread divergence and synchronization restrict the \gls{gpu}'s performance. In contrast, \mempool{} gives each \gls{pe} its instruction stream, making it much more flexible and efficient on irregular workloads.

% Mesh, still shared-memory technically
An alternative way to scale to hundreds of \glspl{pe} is to sacrifice the low-latency interconnect between memory banks and processors and directly connect neighboring cores through a 2D mesh (or similar, e.g., multi-mesh, Ruche, etc.) interconnect. Many multicore \glspl{soc} have been designed following this pattern ~\cite{Olofsson2015, Bell2008, Bohnenstiehl2017, McKeown2017, Taylor2002, Davidson2018}. As an example, RAW~\cite{Taylor2002} connects 16 small cores with private instruction and data memories in a \by{4}{4} grid. The \Glspl{pe} are enhanced with neighbor-to-neighbor communication instructions, and sending data between the \glspl{pe} furthest apart takes only six cycles. Celerity~\cite{Davidson2018} and KiloCore~\cite{Bohnenstiehl2017} implement a similar architecture scaled to 496 or 1000 cores. Replication enables high core counts, but communication latency between distant cores quickly increases to 45 or 64 cycles, respectively, requiring careful dataflow management.

Piton~\cite{McKeown2017} also connects cores in a mesh structure, but instead of using small cores, each of the 25 \glspl{pe} consists of a Linux-capable 64-bit core with private instruction and data caches. TILE64~\cite{Bell2008} and Epiphany-V~\cite{Olofsson2015} connect similar cores, caches, and \glspl{dma} in a 2D mesh but scale the architecture to up to 1024 cores. While all \glspl{pe} still have access to other \gls{pe}'s memory, the \gls{noc} limits the inter-core bandwidth and imposes a high latency. Programming distributed cores requires distributing workloads in a spatially-aware fashion and carefully fitting local data within local memories to reach acceptable performance. Consequently, efficiently programming this class of architectures is challenging, especially for algorithms requiring \glspl{pe} to access a shared pool of data. \mempool{}'s low-latency interconnect greatly simplifies workload distribution and access to shared data, even when such accesses follow irregular and hard-to-predict patterns.

% Mesh, no shared-memory technically
Similarly to KiloCore, Google's Pixel Visual Core~\cite{Redgrave2018} and TPU~\cite{Jouppi2017} connect their \glspl{pe} in a 2D mesh to scale to 2048 or 64K \glspl{pe}, respectively. The \glspl{pe} are only equipped with private register files and communicate with their direct neighbors in a systolic fashion. While these architectures achieve very high performance for their specialized use cases, their rigid interconnect and programming model limit their flexibility.

In summary, \mempool{} is the first architecture to scale the shared-memory cluster with a low-latency interconnect to hundreds of individually programmable \glspl{pe}. In contrast to smaller cluster-based architectures, \mempool{} allows for more parallelism and processing power while providing a larger L1 memory to reduce communication overhead and facilitate latency hiding from/to higher-level memories. We note that there is a physical limit to scaling a single cluster, and for extremely large core counts (thousands), moving to multiple clusters is inevitable. The \mempool{} architecture remains attractive even in a multi-cluster regime, as it reduces the need to partition shared data and greatly eases inter-cluster communication latency hiding with large block transfers.

%%%%%%%%%%%%%%%%%%%%%%%%%%%%%%%%%%%%%%%%%%%%%%%%%%%%%%%%%%%%%%%%%%%%%%%%%%%%%%%
%
% Conclusion
%
%%%%%%%%%%%%%%%%%%%%%%%%%%%%%%%%%%%%%%%%%%%%%%%%%%%%%%%%%%%%%%%%%%%%%%%%%%%%%%%

\section{Conclusion}\label{sec:conclusion}

\mempool{} presents a scalable, shared-L1-memory manycore \riscv{} system. Even when scaling to 256 cores, all cores can access inter-tile or intra-cluster memory banks within at most five cycles of latency thanks to a sophisticated hierarchical \gls{spm} interconnect. Thanks to the capability to program each core independently, the presented cache hierarchy, and the refill interconnect, all cores can be highly utilized, i.e., up to \SI{96}{\percent} in common signal and image processing kernels. Efficient data movement is implemented by a distributed \gls{dma} design integrated into the hierarchical design.

A full 256 core \mempool{} instance is implemented and evaluated in GlobalFoundries' 22FDX technology. It runs at \SI{600}{\mega\hertz} (60 gate delays) in typical operating conditions (TT/\SI{0.80}{\volt}\kern-.1em/\SI{25}{\celsius}) and occupies an area of \SI{12.8}{\milli\meter\squared}. By tailoring the Snitch cores to \mempool{}, and through enhancements like our \gls{dsp} extension and hybrid addressing scheme, \mempool{} achieves very high performance and comes close to an ideally scaled system (e.g., full linear speedup).

Benchmarking on a comprehensive set of kernels shows that \mempool{}'s performance is limited mainly by the inherent load-store architecture and synchronization associated with parallel programming. Architectural stalls contribute only a few percent of speedup loss. Thanks to a Halide framework and  OpenMP runtime integration, implementing complex and irregular parallel kernels is straightforward, as illustrated with a histogram equalization, ray tracing, and \gls{bfs} implementation. In 22FDX technology, a \mempool{} cluster achieves a performance of up to \SI{229}{\giga\ops} reaching an energy efficiency of \SI{180}{\giga\ops\per\watt}. On average, access to shared memory only causes \SI{4}{\percent} stalls, and \mempool{} achieves a per-core \gls{ipc} of 0.96 (over a theoretical bound of 1).

\ifCLASSOPTIONcompsoc{}
  \section*{Acknowledgments}
\else
  \section*{Acknowledgment}
\fi

This work was supported by the ETH Future Computing Laboratory (EFCL), financed by a donation from Huawei Technologies. It also received funding from the HCSCA project \#180625 funded by the Croatian-Swiss Research Programme, and by the European Union's Horizon 2020 research and innovation programme under grant agreement 101070374 (Convolve).

% References
\Urlmuskip=0mu plus 1mu\relax
\def\UrlBreaks{\do\/\do-}
\bibliographystyle{IEEEtran}
\bibliography{bib/mempool}

\begin{thebibliography}{10}
\providecommand{\url}[1]{#1}
\csname url@samestyle\endcsname
\providecommand{\newblock}{\relax}
\providecommand{\bibinfo}[2]{#2}
\providecommand{\BIBentrySTDinterwordspacing}{\spaceskip=0pt\relax}
\providecommand{\BIBentryALTinterwordstretchfactor}{4}
\providecommand{\BIBentryALTinterwordspacing}{\spaceskip=\fontdimen2\font plus
\BIBentryALTinterwordstretchfactor\fontdimen3\font minus \fontdimen4\font\relax}
\providecommand{\BIBforeignlanguage}[2]{{%
\expandafter\ifx\csname l@#1\endcsname\relax
\typeout{** WARNING: IEEEtran.bst: No hyphenation pattern has been}%
\typeout{** loaded for the language `#1'. Using the pattern for}%
\typeout{** the default language instead.}%
\else
\language=\csname l@#1\endcsname
\fi
#2}}
\providecommand{\BIBdecl}{\relax}
\BIBdecl

\bibitem{Muralidhar2022}
R.~Muralidhar, R.~Borovica-Gajic, and R.~Buyya, ``Energy efficient computing systems: Architectures, abstractions and modeling to techniques and standards,'' \emph{ACM Comput. Surv.}, vol.~54, no. 11s, pp. 1--37, Sep. 2022.

\bibitem{Li2010}
H.~Li and N.~Homer, ``A survey of sequence alignment algorithms for next-generation sequencing,'' \emph{Brief. Bioinform.}, vol.~11, no.~5, pp. 473--483, May 2010.

\bibitem{Karlrupp2022}
\BIBentryALTinterwordspacing
K.~Rupp, ``Microprocessor trend data,'' 2022. [Online]. Available: \url{https://github.com/karlrupp/microprocessor-trend-data}
\BIBentrySTDinterwordspacing

\bibitem{NVIDIACorporation2022}
\BIBentryALTinterwordspacing
{NVIDIA Corp.}, ``{NVIDIA} {H100} tensor core {GPU} architecture,'' NVIDIA Corp., Tech. Rep., 2022. [Online]. Available: \url{https://www.nvidia.com/en-us/data-center/h100//}
\BIBentrySTDinterwordspacing

\bibitem{Rocki2020}
K.~Rocki \emph{et~al.}, ``Fast s1/256tencil-code computation on a wafer-scale processor,'' in \emph{Int. Conf. High Perform. Comput. Networking, Storage Anal.}\hskip 1em plus 0.5em minus 0.4em\relax Atlanta, Georgia: IEEE, Oct. 2020, p.~14.

\bibitem{Redgrave2018}
J.~Redgrave, A.~Meixner, N.~Goulding-Hotta, A.~Vasilyev, and O.~Shacham, ``{Pixel Visual Core}: {Google}'s fully programmable image, vision and {AI} processor for mobile devices,'' in \emph{2018 {IEEE} Hot Chips 30 Symp.}, Cupertino, US, Aug. 2018, pp. 1--18.

\bibitem{Hennessy2017}
J.~L. Hennessy and D.~A. Patterson, \emph{Computer Architecture: A Quantitative Approach}, 6th~ed.\hskip 1em plus 0.5em minus 0.4em\relax San Francisco, CA, USA: Morgan Kaufmann, 2017.

\bibitem{Apple2022}
\BIBentryALTinterwordspacing
{Apple Inc.}, ``{Apple} unveils {M1 Ultra}, the world's most powerful chip for a personal computer,'' 2022. [Online]. Available: \url{https://nr.apple.com/d2I1v3s8D5}
\BIBentrySTDinterwordspacing

\bibitem{IntelCorporation2022}
\BIBentryALTinterwordspacing
{Intel Corporation}, ``Intel{\textregistered} core™ i9-12900ks processor,'' 2022. [Online]. Available: \url{https://www.intel.com/content/www/us/en/products/sku/225916/intel-core-i912900ks-processor-30m-cache-up-to-5-50-ghz/specifications.html}
\BIBentrySTDinterwordspacing

\bibitem{Ampere2022}
\BIBentryALTinterwordspacing
{Ampere Computing Corp.}, ``{Ampere}{\textregistered} {Altra}{\textregistered} 64-bit multi-core processor features,'' 2022. [Online]. Available: \url{https://d1o0i0v5q5lp8h.cloudfront.net/ampere/live/assets/documents/Altra_Rev_A1_DS_v1.30_20220728.pdf}
\BIBentrySTDinterwordspacing

\bibitem{GreenWavesTechnologies2021}
\BIBentryALTinterwordspacing
{GreenWaves Technologies SAS}, ``{GAP9} next generation processor for hearables and smart sensors,'' {GreenWaves Technologies SAS}, Tech. Rep., 2021. [Online]. Available: \url{https://greenwaves-technologies.com/wp-content/uploads/2022/06/Product-Brief-GAP9-Sensors-General-V1_14.pdf}
\BIBentrySTDinterwordspacing

\bibitem{DupontdeDinechin2021}
B.~{Dupont de Dinechin}, ``A qualitative approach to many‐core architecture,'' in \emph{Multi-Processor System-on-Chip 1: Architectures}, L.~Andrade and F.~Rousseau, Eds.\hskip 1em plus 0.5em minus 0.4em\relax Hoboken, New Jersey, USA: Wiley, Apr. 2021, ch.~2, pp. 27--51.

\bibitem{Ginosar2016}
R.~Ginosar, P.~Aviely, T.~Israeli, and H.~Meirov, ``{RC64}: High performance rad-hard manycore,'' \emph{IEEE Aerosp. Conf. Proc.}, pp. 2074--2082, Jun. 2016.

\bibitem{Zaruba2021}
F.~Zaruba, F.~Schuiki, and L.~Benini, ``{Manticore}: A 4096-core {RISC-V} chiplet architecture for ultra-efficient floating-point computing,'' \emph{IEEE Micro}, vol.~41, no.~2, pp. 36--42, 2020.

\bibitem{Ditzel2021}
D.~Ditzel \emph{et~al.}, ``Accelerating {ML} recommendation with over a thousand {RISC-V}/tensor processors on {Esperanto}'s {ET-SoC-1} chip,'' in \emph{2021 IEEE Hot Chips 33 Symp.}\hskip 1em plus 0.5em minus 0.4em\relax Palo Alto, California: IEEE, Aug. 2021, pp. 209--220.

\bibitem{Cavalcante2021}
M.~Cavalcante, S.~Riedel, A.~Pullini, and L.~Benini, ``{MemPool}: A shared-{L1} memory many-core cluster with a low-latency interconnect,'' in \emph{2021 Des. Autom. Test Eur. Conf. Exhib.}\hskip 1em plus 0.5em minus 0.4em\relax Grenoble, France: IEEE, 2020, pp. 701--706.

\bibitem{Ragan-Kelley2012}
J.~Ragan-Kelley, A.~Adams, S.~Paris, M.~Levoy, S.~Amarasinghe, and F.~Durand, ``Decoupling algorithms from schedules for easy optimization of image processing pipelines,'' \emph{ACM Trans. Graph.}, vol.~31, no.~4, pp. 1--12, Jul. 2012.

\bibitem{Zaruba2020}
F.~Zaruba, F.~Schuiki, T.~Hoefler, and L.~Benini, ``{Snitch}: A tiny pseudo dual-issue processor for area and energy efficient execution of floating-point intensive workloads,'' \emph{IEEE Trans. Comput.}, vol.~70, no.~11, pp. 1845--1860, Feb. 2021.

\bibitem{Mazzola2021}
S.~Mazzola, S.~Riedel, M.~Cavalcante, L.~Benini, and A.~Macii, ``{ISA} extensions in the {Snitch} processor for signal processing,'' Master's thesis, Politecnico di Torino, Apr. 2021.

\bibitem{Dally2004}
W.~J. Dally and B.~P. Towles, \emph{Principles and Practices of Interconnection Networks}.\hskip 1em plus 0.5em minus 0.4em\relax San Francisco, CA, USA: Morgan Kaufmann, 2004.

\bibitem{Benz2023}
T.~Benz \emph{et~al.}, ``A high-performance, energy-efficient modular dma engine architecture,'' \emph{arXiv Prepr. arXiv2305.05240}, May 2023.

\bibitem{Riedel2021}
S.~Riedel, F.~Schuiki, P.~Scheffler, F.~Zaruba, and L.~Benini, ``Banshee: A fast {LLVM}-based {RISC-V} binary translator,'' in \emph{IEEE/ACM Int. Conf. Comput. Des.}\hskip 1em plus 0.5em minus 0.4em\relax IEEE, Nov. 2021, pp. 1105--1113.

\bibitem{Tagliavini2018}
G.~Tagliavini, D.~Cesarini, and A.~Marongiu, ``Unleashing fine-grained parallelism on embedded many-core accelerators with lightweight {OpenMP} tasking,'' \emph{IEEE Trans. Parallel Distrib. Syst.}, vol.~29, no.~9, pp. 2150--2163, Sep. 2018.

\bibitem{Taylor2002}
M.~B. Taylor \emph{et~al.}, ``The {Raw} microprocessor: A computational fabric for software circuits and general-purpose programs,'' \emph{IEEE Micro}, vol.~22, no.~2, pp. 25--35, Mar. 2002.

\bibitem{Davidson2018}
S.~Davidson \emph{et~al.}, ``The {Celerity} open-source 511-core {RISC-V} tiered accelerator fabric: Fast architectures and design methodologies for fast chips,'' \emph{IEEE Micro}, vol.~38, no.~2, pp. 30--41, Mar. 2018.

\bibitem{Bohnenstiehl2017}
B.~Bohnenstiehl \emph{et~al.}, ``{KiloCore}: A 32-nm 1000-processor computational array,'' \emph{IEEE J. Solid-State Circuits}, vol.~52, no.~4, pp. 891--902, Apr. 2017.

\bibitem{McKeown2017}
M.~McKeown \emph{et~al.}, ``{Piton}: A manycore processor for multitenant clouds,'' \emph{IEEE Micro}, vol.~37, no.~2, pp. 70--80, Mar. 2017.

\bibitem{Bell2008}
S.~Bell \emph{et~al.}, ``{TILE64}™ - processor: A 64-core {SoC} with mesh interconnect,'' in \emph{Dig. Tech. Pap. - IEEE Int. Solid-State Circuits Conf.}, vol.~51.\hskip 1em plus 0.5em minus 0.4em\relax San Francisco, CA, USA: IEEE, 2008, pp. 87--89.

\bibitem{Olofsson2015}
A.~Olofsson, T.~Nordstr{\"{o}}m, and Z.~Ul-Abdin, ``Kickstarting high-performance energy-efficient manycore architectures with {Epiphany},'' in \emph{Conf. Rec. - Asilomar Conf. Signals, Syst. Comput.}\hskip 1em plus 0.5em minus 0.4em\relax IEEE, Apr. 2015, pp. 1719--1726.

\bibitem{Jouppi2017}
N.~P. Jouppi \emph{et~al.}, ``In-datacenter performance analysis of a tensor processing unit,'' in \emph{Proc. - Int. Symp. Comput. Archit.}\hskip 1em plus 0.5em minus 0.4em\relax Toronto, ON, Canada: IEEE, 2017, pp. 1--12.

\end{thebibliography}

% Biography section
\vspace{-5mm}
\begin{IEEEbiography}[{\includegraphics[width=1in,height=1.25in,keepaspectratio,clip]{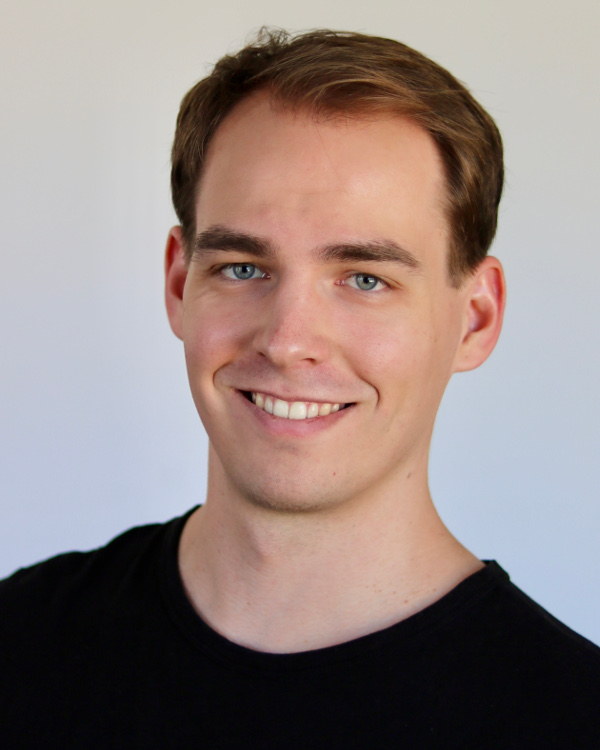}}]{Samuel Riedel}
received the B.Sc. and M.Sc. degrees in Electrical Engineering and Information Technology at ETH Zurich in 2017 and 2019, respectively. He is currently pursuing a Ph.D. degree with the Digital Circuits and Systems group of Luca Benini. His research interests include computer architecture, focusing on manycore systems and their programming model.
\end{IEEEbiography}
\vspace{-5mm}
\begin{IEEEbiography}[{\includegraphics[width=1in,height=1.25in,keepaspectratio,clip]{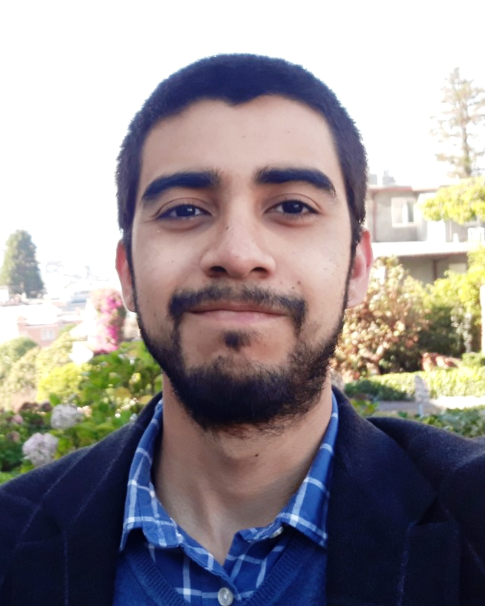}}]{Matheus Cavalcante}
received the M.Sc.\ degree in Integrated Electronic Systems from the Grenoble Institute of Technology (Phelma) in 2018, and is since then pursuing the Ph.D.\ degree at ETH Zurich, with the Digital Circuits and Systems Group of Prof.\ Luca Benini. His current research interests include the design of very-large-scale circuits and high-performance systems, namely vector and manycore architectures, and their co-optimization with emerging VLSI technologies.
\end{IEEEbiography}
\vspace{-5mm}
\begin{IEEEbiography}[{\includegraphics[width=1in,height=1.25in,keepaspectratio,clip]{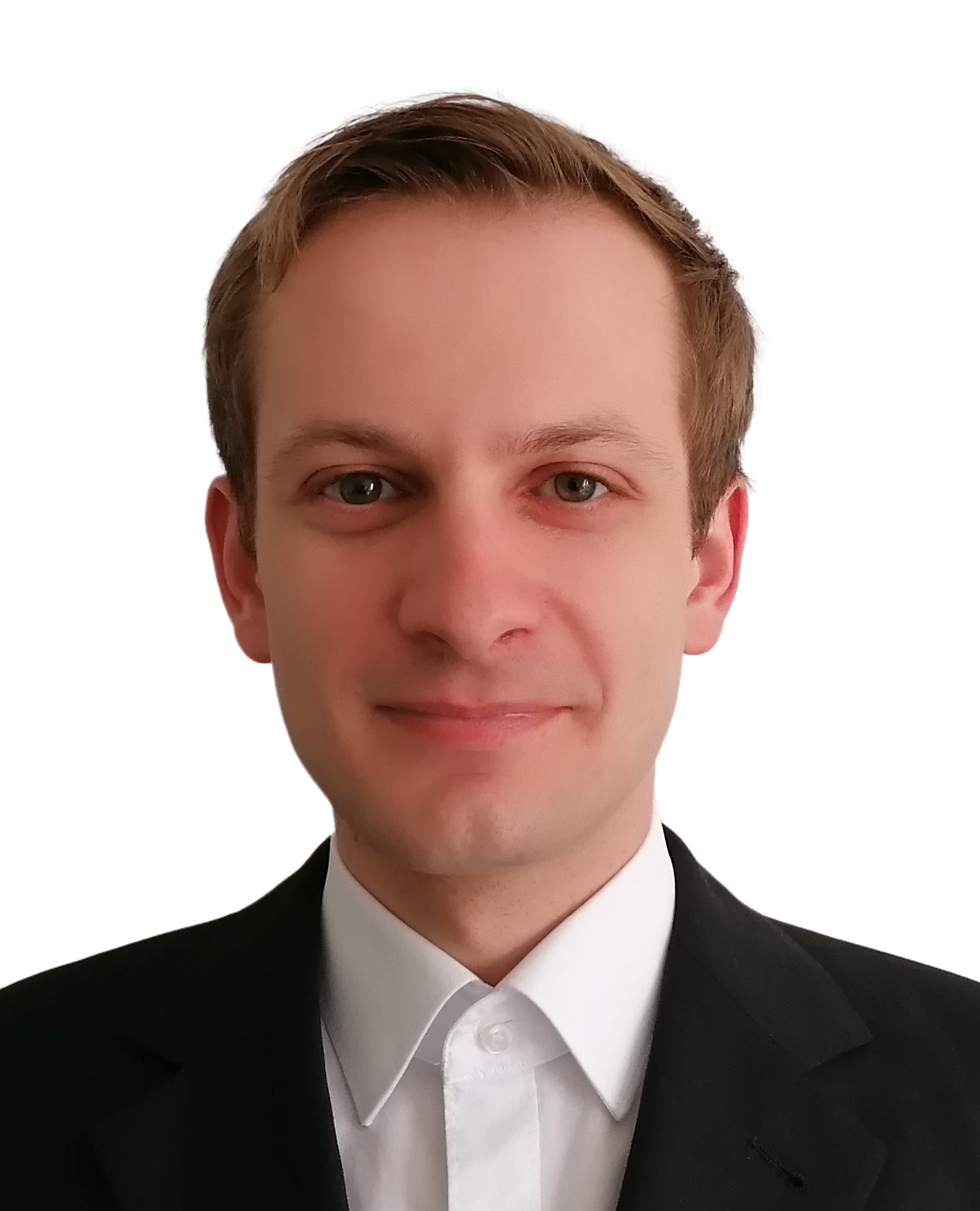}}]{Renzo Andri}
received the B.Sc., M.Sc., and Ph.D. degrees in Electrical Engineering and Information Technology at ETH Zurich in 2013, 2015, and 2020, respectively. His research focuses on energy-efficient machine learning acceleration from system-level design to full-custom IC design. In 2019, he won the IEEE TCAD Donald O. Pederson Award.
\end{IEEEbiography}
\vspace{-5mm}
\begin{IEEEbiography}[{\includegraphics[width=1in,height=1.25in,keepaspectratio,clip]{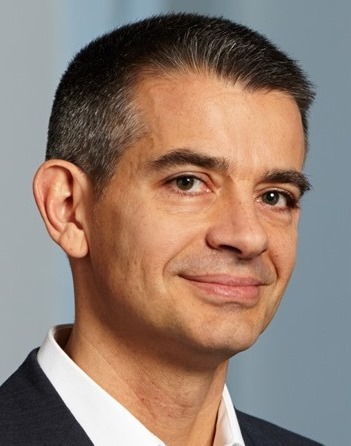}}]{Luca Benini}
is the Chair of Digital Circuits and Systems at  ETH Zürich and a Full Professor at the University of Bologna.
He has served as Chief Architect for the Platform2012 in STMicroelectronics, Grenoble.
Dr.\ Benini's research interests are in energy-efficient systems and multi-core SoC design.
He is a Fellow of the ACM and a member of the Academia Europaea.
\end{IEEEbiography}

\end{document}